\ifdefined\pdfoutput\pdfoutput=1\fi
\documentclass[reprint,amsmath,amssymb,aps,nofootinbib,superscriptaddress,showkeys]{revtex4-2}

\usepackage{amsmath,amssymb,amsfonts}
\usepackage{graphicx,array,dcolumn}
\usepackage{soul}

\usepackage{amsthm}
\usepackage[figuresright]{rotating}
\usepackage{algorithm,algorithmicx,algpseudocode}
\usepackage{listings}
\usepackage{xcolor}
\usepackage{pifont}
\usepackage{hyperref}
\usepackage{bm,gensymb,ulem,bbm,animate,subcaption,mathtools}

\makeatletter

\newcommand{\ket}[1]{\vert #1 \rangle}

\newcommand{\eV}{{\text{eV}}}
\newcommand{\eVA}{{\text{eV}\cdot \text{\AA}}}
\newcommand{\bk}{\mathbf{k}}
\newcommand{\bq}{\mathbf{q}}

\newcommand{\btk}{\widetilde{\mathbf{k}}}
\newcommand{\btq}{\widetilde{\mathbf{q}}}
\newcommand{\br}{\mathbf{r}}

\newcommand{\I}{\uppercase\expandafter{\romannumeral 1}}
\newcommand{\II}{\uppercase\expandafter{\romannumeral 2}}
\newcommand{\III}{\uppercase\expandafter{\romannumeral 3}}
\newcommand{\IV}{\uppercase\expandafter{\romannumeral 4}}
\newcommand{\V}{\uppercase\expandafter{\romannumeral 5}}
\newcommand{\VI}{\uppercase\expandafter{\romannumeral 6}}
\newcommand{\VII}{\uppercase\expandafter{\romannumeral 7}}

\newcommand{\bfa}{\mathbf{a}}
\newcommand{\bfb}{\mathbf{b}}
\newcommand{\p}{\mathbf{p}}

\newcommand{\bfk}{\mathbf{k}}
\newcommand{\G}{\mathbf{G}}

\newcommand{\q}{\mathbf{q}}

\newcommand{\hc}{\hat{c}}
\newcommand{\hcd}{\hat{c}^{\dagger}}

\newcommand{\R}{\mathbf{R}}
\newcommand{\B}{\mathbf{B}}

\def\q{\mathbf{q}}
\def\uns{\ifmmode\,\else$\,$\fi}

\makeatother

\begin{document}

\title{Fermi lune and non-reciprocal transport in rhombohedral multilayer graphene}

\author{Min Li,$^{1,*}$ Qingxin Li,$^{2,*}$ Xin Lu,$^{1}$ Hua Fan,$^{3}$ Kenji Watanabe,$^{4}$ Takashi Taniguchi,$^{5}$ Yue Zhao,$^{3}$ Xin-Cheng Xie,$^{6,7}$ Lei Wang,$^{2,\dagger}$ and Jianpeng Liu$^{1,8,\dagger}$}
\noaffiliation

\affiliation{State Key Laboratory of Quantum Functional Materials, School of Physical Science and Technology, ShanghaiTech Laboratory for Topological Physics, ShanghaiTech University, Shanghai 201210, China}
\affiliation{National Laboratory of Solid-State Microstructures, Collaborative Innovation Center of Advanced Microstructures, School of Physics, Nanjing University, Nanjing, China}
\affiliation{Department of Physics, State Key Laboratory of Quantum Functional Materials, Guangdong Basic Research Center of Excellence for Quantum Science, Southern University of Science and Technology, Shenzhen 518055, China}
\affiliation{Research Center for Electronic and Optical Materials, National Institute for Materials Science, 1-1 Namiki, Tsukuba 305-0044, Japan}
\affiliation{Research Center for Materials Nanoarchitectonics, National Institute for Materials Science, 1-1 Namiki, Tsukuba 305-0044, Japan}
\affiliation{Interdisciplinary Center for Theoretical Physics and Information Sciences, Fudan University, Shanghai 200433, China}
\affiliation{International Center for Quantum Materials, School of Physics, Peking University, Beijing 100871, China}
\affiliation{Liaoning Academy of Materials, Shenyang 110167, China}

\renewcommand{\thefootnote}{\fnsymbol{footnote}}
\setcounter{footnote}{0}
\footnotetext{These authors contributed equally to this work.}
\footnotetext{Corresponding authors. Email: leiwang@nju.edu.cn; liujp@shanghaitech.edu.cn.}

\begin{abstract}
We develop a microscopic theory of Fermi-lune metals in rhombohedral multilayer graphene and investigate their non-reciprocal longitudinal transport.  We show that tuning the layer number and displacement field drives Lifshitz transitions from circular to annular non-interacting Fermi surfaces accompanied by redistributions of electronic weight among layers. Combining renormalization-group corrections with unrestricted Hartree--Fock calculations, we further demonstrate how  Coulomb interactions reconstruct the non-interacting annular Fermi surface into crescent-shaped Fermi lune through spontaneous breaking of time-reversal, mirror and rotation symmetries.
We determine competing mean-field phases for four to nine layers and distinguish two types of transdimensional orbital magnetic orders. Beyond the  Fermi-lune state that exhibits transdimensional anomalous Hall effect reported recently, we also identify a mirror-preserving Fermi-lune phase with spontaneous in-plane orbital magnetization but a symmetry-forbidden anomalous Hall response. The maximum calculated in-plane orbital moment increases approximately linearly with layer number. The asymmetric Fermi-velocity distributions of these states also lead to intrinsic non-reciprocal longitudinal transport. Measurements presented here in a nine-layer device show a non-reciprocal resistance ratio reaching $23\%$ at a current magnitude of $17\,\mathrm{nA}$. These results elucidate the formation mechanism, competing symmetries, and thickness dependence of Fermi-lune states, and connect their electronic structure to a longitudinal-transport signature beyond the previously reported anomalous Hall phenomenology.
\end{abstract}


\maketitle

\section{Introduction}
Metallic systems are usually characterized by the presence of robust Fermi surfaces, which make dominant contributions to the low-energy physical properties. Symmetry and geometry of the Fermi surface, along with the corresponding gapless ground state, play a pivotal role in determining the transport properties \cite{kaganov1979electron,springford2011electrons}. It is well known that anomalous Hall effect arises in metals breaking time-reversal symmetry \cite{nagaosa-ahe-rmp-2010}. Nonlinear Hall effect arises due to the breaking of inversion symmetry \cite{nlhe-nrp21}. A Fermi surface breaking both time-reversal and inversion symmetries in the absence of external electromagnetic fields would exhibit both non-reciprocal charge transport and anomalous Hall effect \cite{tokura-natnano20,li-nm24}, which is seldom reported. 
Evidently, Fermi surfaces with reduced symmetry are closely linked with exotic electromagnetic responses. The discovery of new class of Fermi surface structure may reveal fundamentally new physical phenomena.

Rhombohedral multilayer graphene (RMG) provides a tunable setting for studying novel metallic states. Its carrier density, vertical displacement field, and layer number control both the low-energy dispersion and the layer distribution of electronic wavefunctions, which play pivotal roles in the low-energy transport properties. Earlier studies established spin- and valley-polarized metals in graphene multilayers~\cite{zhou-halfmetal_R3G-nature-2021,liu-symbreakR4G-natnano-2024} and explored momentum-polarized and current-carrying orders~\cite{levitov-prb23,li-arxiv23,prb15_mac_currentstate,prb23_fs_trilayer}. Here we focus on the transdimensional regime, where the system thickness $d$ satisfies $d_0 \ll d \lesssim l_{\perp}$, with $d_0$ the atomic-layer thickness and $l_{\perp}$ the out-of-plane coherence length, such that coherent interlayer orbital motion persists across the sample and plays an essential role in the low-energy physics. In this regime, Fermi-surface reconstruction can modify orbital motions across the sample thickness as well as within the graphene plane. A new type of anomalous Hall effect has been discovered experimentally in RMG in the transdimensional regime, termed ``transdimensional anomalous Hall effect" (TDAHE) \cite{comment_exp}, where the Hall-resistance hysteresis is controlled by both in-plane and out-of-plane magnetic fields.  The  TDAHE is associated with the emergence of spontaneous out-of-plane and in-plane orbital current loops, contributing to both in-plane and out-of-plane orbital magnetization components.  Remarkably, such ``transdimensional orbital magnetic" states exhibit an intriguing crescent-shaped Fermi surface, termed ``Fermi lune" here in this work.  

Ref.~\onlinecite{comment_exp} includes a selected theoretical result of the crescent-shaped Fermi surface to theoretically explain the TDAHE. However, the microscopic formation mechanism, the competing symmetry-breaking states, the dependence on layer numbers and displacement fields and other new transport properties, are all open questions.  In this work, we establish a microscopic picture for the formation and physical consequences of the Fermi-lune state. We first show that increasing either the layer number or the displacement field drives a Lifshitz transition of the non-interacting Fermi surface from a circular pocket to an annulus, accompanied by a redistribution of the electronic wavefunctions from the outermost layer toward a more extended multilayer profile. Incorporating electron-electron Coulomb interactions through unrestricted Hartree--Fock calculations assisted by renormalization-group corrections, we find that the annular Fermi surface is generically unstable toward spontaneous momentum-space symmetry breaking, giving rise to crescent-shaped Fermi lunes and transdimensional orbital magnetic states. We identify distinct competing Fermi-lune states with different symmetry properties, map out their evolution with carrier density, displacement field, and layer number, and elucidate their associated in-plane and out-of-plane orbital responses. Most importantly, the strongly asymmetric Fermi velocities of the Fermi lune naturally generate intrinsic non-reciprocal longitudinal transport, which we observe experimentally in  nine-layer RMG together with TDAHE. These results establish the Fermi lune as a distinct interaction-driven metallic state and reveal transport phenomena beyond the TDAHE reported in Ref.~\onlinecite{comment_exp}.

\begin{figure*}[hbtp!]
	\centering 
	\includegraphics[width=\textwidth]{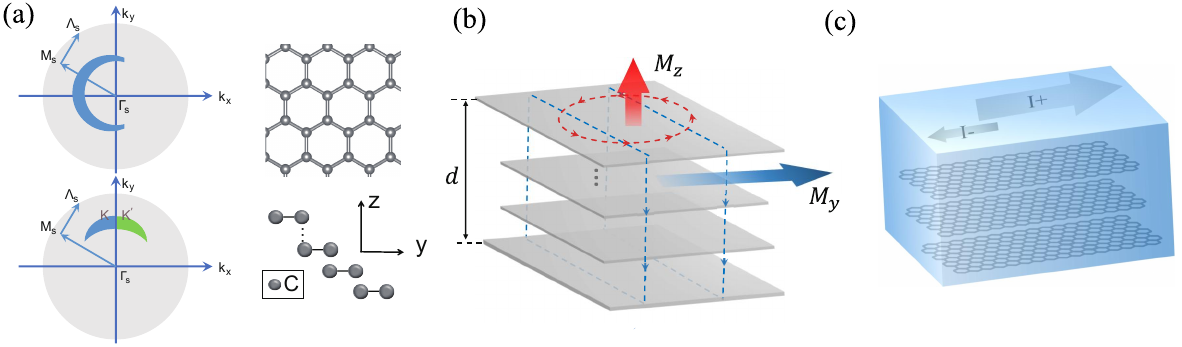}
	\caption{\textbf{Schematic illustrations of Fermi lune and transdimensional orbital magnetism.} 
		(a) Left: schematic illustration of two types of Fermi lunes. The upper Fermi lune, corresponding to the TOM$_y$ state, breaks $\mathcal{T}$, $C_3$, and $\mathcal{M}_y$, while preserving $\mathcal{M}_y\mathcal{T}$. Here $\mathcal{M}_y$ denotes the vertical mirror plane containing the crystalline $y$ axis, namely $\mathcal{M}_y:(x,y,z)\rightarrow(-x,y,z)$. The lower Fermi lune, corresponding to the TOM$_x$ state, breaks $\mathcal{T}$, $C_3$, and $\mathcal{M}_y\mathcal{T}$, while preserving $\mathcal{M}_y$. Right: illustration of the lattice structure of rhombohedral multilayer graphene. 
		(b) Schematic illustration of transdimensional orbital magnetism, which is characterized by both in-plane and out-of-plane orbital magnetizations. 
		(c) Illustration of non-reciprocal longitudinal transport.}
	\label{fig1}
\end{figure*}

\section{Results}
\subsection{Non-interacting Fermi surfaces and Lifshitz transitions}

\begin{figure*}[hbtp!]
	\centering
    \includegraphics[width=\textwidth]{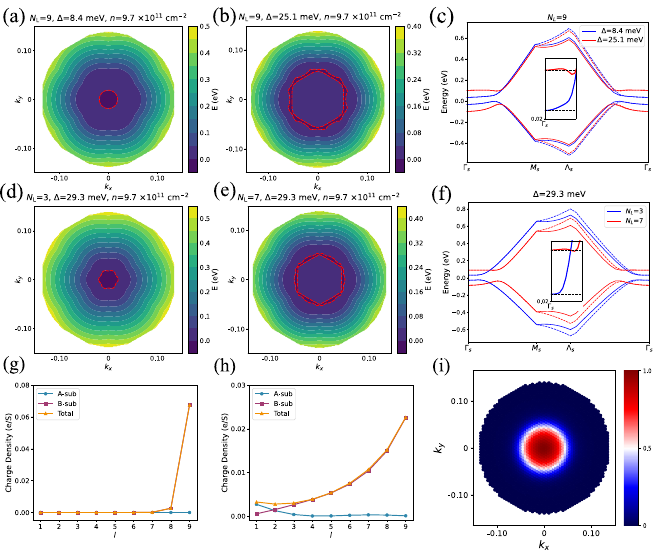}
	\caption{\textbf{Non-interacting Fermi surfaces and Lifshitz transitions.} 
		Non-interacting Fermi surfaces of 9-layer RMG at a carrier density of $n=9.7\times 10^{11}\,\text{cm}^{-2}$ for (a) interlayer electrostatic energy drop $\Delta=8.4\,$meV and (b) $\Delta=25.1\,$meV. 
		(c) Non-interacting band structures of 9-layer RMG, where the blue and red lines represent the energy bands with $\Delta=8.4\,$meV and $\Delta=25.1\,$meV, respectively. 
		Non-interacting Fermi surfaces with fixed $\Delta=29.3\,$meV and $n=9.7\times 10^{11}\,\text{cm}^{-2}$ for (d) 3-layer RMG and (e) 7-layer RMG. 
		(f) Non-interacting band structures of RMG with $\Delta=29.3\,$meV, where the blue and red lines correspond to the energy bands of 3-layer and 7-layer RMG, respectively. 
		(g),(h) Layer-resolved electron density distributions of 9-layer RMG at a carrier density of $n=9.7\times 10^{11}\,\text{cm}^{-2}$ for (g) $\Delta=8.4\,\text{meV}$ and (h) $\Delta=25.1\,\text{meV}$. 
		(i) Top-layer contributions to the wavefunction of the lowest conduction band in $\mathbf{k}$-space, showing that only the $\mathbf{k}$ points near the Dirac point (the origin) are dominated by the top layer.}
	\label{fig2}
\end{figure*}

As discussed above, the Fermi-lune state emerges in RMG in the transdimensional regime. Therefore, prior to analyzing the core mechanisms in achieving such state, we first estimate the vertical mean free path $l_{\perp}$ for RMG, which defines the characteristic thickness of the transdimensional regime for layered van der Waals materials. By combining the in-plane and out-of-plane conductivity data of bulk graphite \cite{graphite-pr53} with the in-plane resistivity measurements from a 9-layer rhombohedral graphene device, using Drude formula we roughly estimate $l_{\perp}\sim 2\,$nm \cite{comment_exp}. Assuming $l_{\perp}$ remains applicable to RMG with varying number of layers ($N_{\text{L}}$), it follows that systems with $N_{\text{L}}\gtrapprox 5-6$ already enter the transdimensional regime. Within this regime, out-of-plane orbital motions exhibit non-negligible contributions while retaining quantum coherence. As $N_{\text{L}}$ increases further, electrons gradually lose coherence in out-of-plane orbital dynamics, eventually transitioning to a fully three-dimensional regime.

Therefore, in this work we explore the properties of slightly charge-doped RMG with the number of layers $4\leq N_{\rm{L}}\leq 9$.
We first study the non-interacting electronic structures with varying interlayer electrostatic potential energy drop $\Delta$ (induced by a vertical displacement electric field) and electron carrier density $n$. The single-particle properties of low-energy electrons of RMG can be properly described by a continuum model as detailed in Supplementary Materials \cite{supp_info}. In general, the non-interacting Fermi surfaces of slightly electron-doped RMG can be categorized into two groups. In the case of weak $\Delta$ and small $N_{\text{L}}$, the Fermi surface is characterized by a simple circular electron pocket centered at the Dirac point of each valley ($\mathbf{K}$ or $\mathbf{K}'$ valley); while for large $\Delta$ and/or large $N_{\text{L}}$, it consists of a double-ring shaped, or an annular shaped Fermi surface surrounding the Dirac point. As a demonstration, in Fig.~\ref{fig2}(a) and (b) we present the non-interacting Fermi surfaces (marked by red dashed lines) of 9-layer RMG at fixed carrier density $n=9.7\times 10^{11}\,\text{cm}^{-2}$, for $\Delta=8.4\,$meV and $\Delta=25.1\,$meV, respectively. Clearly, with the increase of $\Delta$, the Fermi surface undergoes a Lifshitz transition from a circle to an annulus structure. Fig.~\ref{fig2}(c) presents the corresponding band structures. The conduction band minimum evolves from the Dirac point ($\Gamma_s$) for $\Delta=8.4\,$meV (blue line) to somewhere between $\Gamma_s$ and $M_s$ (red line), leading to the annular shaped Fermi surface under slight electron doping. Similarly, fixing $\Delta=29.3\,$meV and $n=9.7\times 10^{11}\,\text{cm}^{-2}$, the non-interacting Fermi surface topology also changes from circle to annulus as the number of layers $N_{\text{L}}$ increases from 3 to 7, as shown in Fig.~\ref{fig2}(d) and (e), where the Fermi energy contours are marked by red dashed lines. The corresponding band structures are shown in Fig.~\ref{fig2}(f).  

Moreover, in the case of circular Fermi surface, the conduction band electrons are exponentially localized at the top layer as shown in Fig.~\ref{fig2}(g) for the 9-layer system with $\Delta=8.4\,$meV and $n=9.7\times 10^{11}\,\text{cm}^{-2}$. This is because the wavefunctions of the lowest conduction band around the Dirac point is mostly dominated by the top layer (see Fig.~\ref{fig2}(i)), and these top-layer dominated $\mathbf{k}$-space region is occupied by electrons for the case of circular Fermi surface. In contrast, when the Lifshitz transition occurs, the electrons populate a thin annulus surrounding the Dirac point, and the contributions from the top layer states are significantly reduced. This makes the lowest-conduction-band's electrons no longer exponentially localized at the top layer, but become extendedly distributed among all layers though the top layer still has the largest contribution, as shown in Fig.~\ref{fig2}(h) for the 9-layer system with $\Delta=29.3\,$meV and $n=9.7\times 10^{11}\,\text{cm}^{-2}$. 
We will see shortly that the change of Fermi surface topology and the different layer distributions of electrons give rise to fundamentally different spontaneous symmetry-breaking states under small carrier dopings once electron-electron interactions are taken into account. It turns out that TOM state with Fermi-lune structure mostly emerges from the annular shaped non-interacting Fermi surface with more extended layer distributions of electrons.

\subsection{Fermi lune and transdimensional orbital magnetism}

\begin{figure*}[hbtp!]
	\centering 
	\includegraphics[width=\textwidth]{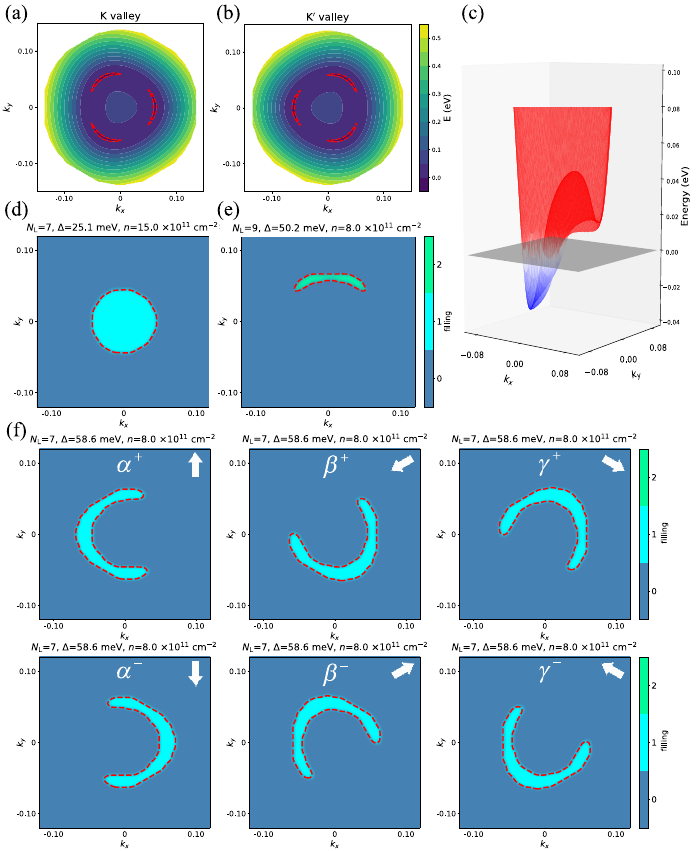}
	\caption{\textbf{Hartree-Fock electronic structures and magnetic domains of Fermi-lune states.}
		(a), (b) Non-interacting Fermi surfaces in the $K$ and $K^\prime$ valleys, showing that each valley hosts three lune-shaped Fermi pockets under an interlayer electric potential drop $\Delta=58.6\,\text{meV}$ and a carrier density of $n=9.7\times 10^{11}\,\text{cm}^{-2}$. (c) Interacting band structures of the TOM$_y$ state, where the blue and red colors indicate occupied and empty energy bands, respectively. The gray horizontal plane set to 0 marks the Fermi energy.  Fermi surfaces of (d) SVP and (e) TOM$_x$ states. (f) The six magnetic domains of the TOM$_y$ state in 7-layer RMG with $\Delta=58.6\,\text{meV}$ and $n=8.0\times 10^{11}\,\text{cm}^{-2}$. The thick white arrows denote the directions of the in-plane orbital magnetizations. The color coding in (d), (e), and (f) indicates the electron filling number in $\mathbf{k}$-space.  $k_x$, $k_y$ are in unit of \AA$^{-1}$.}
	\label{fig3}
\end{figure*}

We further include the long-range electron-electron Coulomb interactions which play the dominant role in slightly doped RMG \cite{supp_info}.
\begin{equation}
	\hat{V}^{\rm{intra}}
	=\frac{1}{2S}\sum_{\lambda\lambda'}\sum_{\bfk \bfk ' \q} V^{\text{dg}}(\q)\,
	\hat{c}^{\dagger}_{\lambda}(\bfk+\q) \hat{c}^{\dagger}_{\lambda'}(\bfk ' - \q) \hat{c}_{\lambda '}(\bfk ')\hat{c}_{\lambda}(\bfk)\;,
	\label{eq:Vintra}
\end{equation}
where $V^{\text{dg}}(\q)=e^2\tanh(|\mathbf{q}|d_s)/(\,2 \epsilon_{\textrm{BN}}\varepsilon_0 |\mathbf{q}|\,)$ is the Fourier transform of double-gate screened Coulomb interaction at wavevector $\q$, with $d_s=40\,$nm, $\epsilon_{\textrm{BN}}=4$ denoting the relative dielectric constant of hexagonal BN (hBN) substrate.  $\lambda=(\mu, \sigma, \alpha)$ refer to the valley, spin and layer/sublattice degrees of freedom, respectively. Electron-electron interactions may drive RMG system to different types of symmetry-breaking states under slight carrier dopings, including flavor-polarized metallic phase \cite{zhou-halfmetal_R3G-nature-2021,delabarrera-isospinbGr-natphys-2022,zhou-isospinSCbGr-science-2022,han-symbreakR5G-natnano-2024,liu-symbreakR4G-natnano-2024}, momentum polarized state (in bilayer and trilayer graphene) \cite{levitov-prb23,li-arxiv23,prb15_mac_currentstate,prb23_fs_trilayer}, and superconductivity \cite{zhou-trilayer-nature21,zhou-blg-science22,zhangyr-blg-nature23,li-blg-nature24,han-chiralsc-nature25}. However, the interaction-driven instabilities of the annular shaped Fermi surface for RMG in the transdimensional regime has rarely been discussed.

\begin{table*}
	\caption{\textbf{Symmetry and physical properties of different symmetry-breaking states} TOM$_y$, TOM$_x$, and SVP are distinguished by their symmetry properties and associated observables. $M_{a}$ ($a=x, y, z$) refers to orbital magnetization in $a$ direction, $\sigma_{yx}$ denotes anomalous Hall conductivity, and $\delta R_{xx}=R^{+}_{xx}-R^{-}_{xx}$ characterizes non-reciprocal longitudinal transport behavior.}
	\label{table:range}
	\centering
	\begin{tabular}{c | c| c| c| c || c| c| c| c|c}
		\hline
		& $\mathcal{T}$ & $C_3$     & $\mathcal{M}_y$  &  $\mathcal{M}_y\mathcal{T}$  & $M_z$     &  $M_y$  & $M_x$ &   $\sigma_{yx}$  & $\delta R_{xx}$  \\		
		\hline
		TOM$_y$  &  $\times$    &  $\times$  &  $\times$        & \checkmark              & \checkmark &  \checkmark  &  $\times$  & \checkmark  & \checkmark \\
		\hline
		TOM$_x$  & $\times$      & $\times$ & \checkmark       &  $\times$                & $\times$  & $\times$   &\checkmark  & $\times$   & \checkmark  \\ 	
		\hline
		SVP      & $\times$    &  \checkmark & $\times$ & $\times$   & \checkmark & $\times$  & $\times$  & \checkmark   & \checkmark \\
		\hline 
	\end{tabular}
\end{table*}

To faithfully capture the Coulomb interaction effects, we set up a low-energy window $E_C^{*}\sim 0.3\,$eV within which $e$-$e$ interactions are treated non-perturbatively using self-consistent Hartree-Fock approach, allowing for the spontaneous breaking of time-reversal, spin rotational and point-group symmetries \cite{supp_info}. 
The occupied electrons outside $E_C^*$ would significantly influence the low-energy properties through long-range Coulomb interactions, which can be treated using perturbative renormalization group approach \cite{kang-rg-prl20,guo-prb24}. As a result, the renormalized low-energy continuum model parameters generally acquire logarithmic enhancement \cite{supp_info}. With the renormalized low-energy single-particle model, we further consider interactions between low-energy electrons, which are obtained by projecting $V^{\rm{intra}}$ onto the low-energy wavefunctions within $E_C^*$. Then, we perform fully unrestricted Hartree-Fock calculations for slightly electron doped RMG with varying $N_{\text{L}}$, $\Delta$ and $n$.   On the one hand, if one starts from a circular non-interacting Fermi surface centered at the Dirac point, the system generally stays in a metallic phase that spontaneous lifts the valley and/or spin degeneracy, yet still maintains the original Fermi-surface topology \cite{supp_info}. We call such metallic state as spin-valley-polarized (SVP) state, which is also known as the ``quarter metal" or ``half-metal" in literatures \cite{zhou-halfmetal_R3G-nature-2021,liu-symbreakR4G-natnano-2024}, depending on how much the spin/valley degeneracy is lifted.  On the other hand, if one starts from an annular shaped Fermi surface, the system would mostly stay in the Fermi-lune state driven by $e$-$e$ interactions, concomitant with the emergence of transdimensional orbital magnetic properties.  

In general, there are two types of TOM states characterized by orthogonal Fermi lune structures, as schematically shown in Fig.~\ref{fig1}(a). One of them spontaneously breaks time-reversal ($\mathcal{T}$), $C_{3}$ and vertical mirror ($\mathcal{M}_y$) symmetries of RMG, while preserves the combined $\mathcal{M}_y\mathcal{T}$ symmetry. This state is characterized by a Fermi lune bisected by crystalline $x$ axis (see left upper panel of Fig.~\ref{fig1}(a)), and possesses in-plane orbital magnetization along the crystalline $y$ axis (and the other two equivalent crystalline axes connected by $C_3$ rotations), dubbed as ``TOM$_y$" state. A three-dimensional plot of the band structures of TOM$_y$ state obtained from Hartree-Fock calculations is shown in Fig.~\ref{fig3}(c), where the blue and red colors indicate energy bands that are occupied and empty, respectively. The occupied bands form an electron pocket with its Fermi energy contour in lune shape. This Fermi lune structure is inherited from the non-interacting Fermi surface of RMG. Specifically, the annular shaped Fermi surface (Fig.~\ref{fig2}(b) and (e)) is contributed by two valleys ($K$ valley and $K'$ valley). If only inspecting the non-interacting Fermi surface of one valley, it consists of three lune-shaped pockets that are related to each other by threefold rotations, as clearly shown in Fig.~\ref{fig3}(a) and (b) for $K$ and $K'$ valleys, respectively, for the case of 7-layer rhombohedral graphene with $\Delta=58.6\,$meV and $n=8.0\times 10^{11}\,$cm$^{-2}$.  Combining the six pieces of lunes from the two valleys, they are connected and form an annulus. This characteristic is an intrinsic consequence of the lattice structure and symmetry of rhombohedral graphene. In the TOM$_y$ state, long-range Coulomb interactions tend to drive the electrons to populate two of the three pieces of lunes from one valley forming a crescent-shaped Fermi contour, which spontaneously breaks time-reversal, $C_{3}$ and $\mathcal{M}_y$ symmetries.

As a result, there are six magnetic domains for the TOM$_y$ state, labelled as $\alpha^{\pm}$, $\beta^{\pm}$ and $\gamma^{\pm}$, as shown in Fig.~\ref{fig3}(f). These six energetically equivalent magnetic domains are connected to each other by $C_3$ rotations, as well as $\mathcal{T}$ and $\mathcal{M}_y$ operations. For example, the $\alpha^{+}$, $\beta^{+}$ and $\gamma^{+}$ domains transform to each other under $C_3$ rotations; the $\alpha^{+}$ and $\alpha^{-}$ are connected to each other either by $\mathcal{T}$ or $\mathcal{M}_y$ operation; while $\beta^{+}$ and $\gamma^{-}$ can transform to each other by $\mathcal{M}_y$ operation, etc. The six domains contribute to in-plane orbital magnetizations along the three equivalent crystalline axes as shown by the thick white arrows in Fig.~\ref{fig3}(f), and the calculated in-plane orbital magnetization is on the order of $1\rm{-}10\,\mu_{\rm{B}}$ per electron for RMG with the number of layers $5\leq N_{\rm{L}}\leq 9$. 
By virtue of the $C_3$, $\mathcal{T}$ and $\mathcal{M}_y$ symmetry breakings, the TOM$_y$ state exhibits transdimensional AHE that couples to both in-plane and out-of-plane orbital magnetizations. Consequently, such AHE exhibits hysteretic response to both in-plane and out-of-plane magnetic fields \cite{comment_exp}. 

The other TOM state spontaneously breaks $C_3$, $\mathcal{T}$ and the combined $\mathcal{M}_y\,\mathcal{T}$ symmetries, while preserves $\mathcal{M}_y$ symmetry. It is characterized by a Fermi lune bisected by the crystalline $y$ axis (and the other two equivalent crystalline axes connected by $C_3$ rotations as shown in Fig.~\ref{fig3}(e), and possesses in-plane orbital magnetization along the crystalline $x$ direction (and the other two equivalent directions connected by $C_3$ rotations), dubbed as ``TOM$_x$" state. This state does not possess any AHE due to the presence of $\mathcal{M}_y$ symmetry, which kills out-of-plane orbital magnetization and anomalous Hall conductivity. The Fermi lune associated with TOM$_x$ state possesses a momentum dependent valley polarization. As shown in the lower left panel of Fig.~\ref{fig1}(a) , electrons in the TOM$_x$ state are polarized to $K$ ($K'$) valley on the left (right) side of the $k_y$ axis passing through the Dirac point, as enforced by $\mathcal{M}_y$ symmetry.  Nevertheless, one would expect that the TOM$_x$ state still possesses non-reciprocal longitudinal transport property due to the pronounced asymmetry between the forward-moving and backward-moving Fermi velocities associated with the Fermi-lune structure.  The TOM$_x$ state can be interpreted as electrons occupying two lune sectors from opposite valleys: one from the $K$ and the other from $K'$. These two sectors are connected by $\mathcal{M}_y$ operation, thereby preserving $\mathcal{M}_y$ symmetry. Similar to the case of TOM$_x$ state, there are also six energetically equivalent magnetic domains, with six orientations of lunes and in-plane magnetization directions. Notably, both TOM$_y$ and TOM$_x$ states are characterized by spontaneous out-of-plane orbital current loops. The corresponding ground-state current circulates within the $x-z$ and $y-z$ planes for TOM$_y$ and TOM$_x$ states, respectively. The detailed, layer-dependent ground-state current distributions are presented in Supplementary Materials.

In Table~\ref{table:range} we enumerate the symmetries and physical properties of the three types of symmetry-breaking states emerging from slightly carrier-doped RMG of transdimensional thickness. The symmetries and the physical properties of the TOM$_y$ and TOM$_x$ states are listed in the first two rows of Table~\ref{table:range}. In particular, the TOM$_y$ state exhibits both transdimensional AHE and non-reciprocal longitudinal transport behavior; while only the latter is present for the TOM$_x$ state due to the presence of $\mathcal{M}_y$ mirror symmetry.  The symmetry and physical properties of the SVP state are given in the last row of Table~\ref{table:range}. While the SVP state breaks both $\mathcal{T}$ and $C_{2z}$ symmetries, allowing in principle for non-reciprocal longitudinal transport, the experimental signal is expected to be weak. This is because the nearly circular Fermi surface, which exhibits only weak trigonal warping (Fig.~\ref{fig3}(d)), results in a slight asymmetry between the Fermi velocities of forward- and backward-moving carriers. We note that the TOM$_x$ state is mostly spin degenerate, while the TOM$_y$ state may or may not have spin degeneracy depending on detailed properties of the system such as $\Delta$, $n$ and number of layers.  Here we do not discuss the spin properties of the TOM states in depth, since the spin degrees of freedom do not contribute to orbital transport properties due to the negligible spin-orbit coupling effects. 

\subsection{Giant non-reciprocal longitudinal transport}
\begin{figure*}[hbtp!]
	\includegraphics[width=\textwidth]{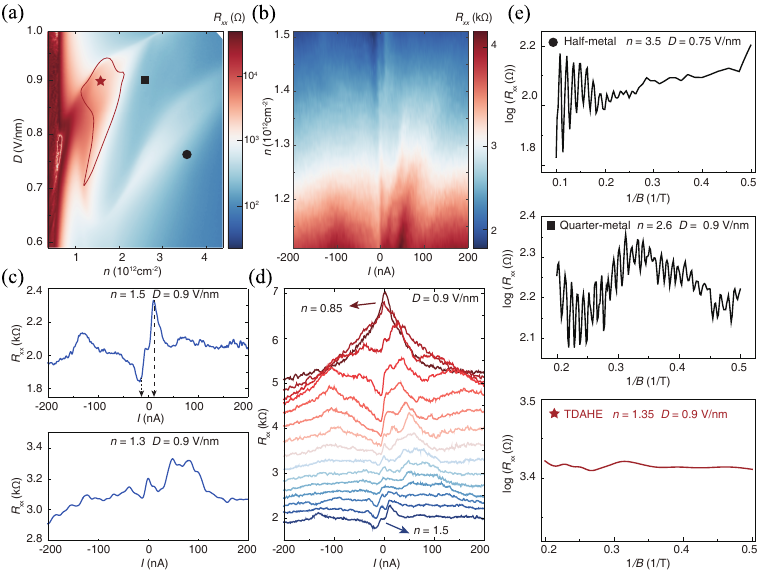}
	\caption{\textbf{Giant non-reciprocal longitudinal transport.} 
		(a) Longitudinal resistance $R_{xx}$ measured in a 9-layer RMG device at temperature 14\,mK, plotted as a function of $D$ and $n$ in logarithmic scale. The solid red line marks the phase boundary within which there is transdimensional AHE. (b) $R_{xx}$ measured for a 9-layer RMG device, as a function of carrier density $n$ and applied current $I$, under fixed $D=0.9\,$V/nm. (c) Line plots of $R_{xx}$ as a function of $I$, upper panel: for $D=0.9\,$V/nm and $n=1.5\times 10^{12}\,\text{cm}^{-2}$, and lower panel:  for $D=0.9\,$V/nm and $n=1.3\times 10^{12}\,\text{cm}^{-2}$. (d) Line plots of $R_{xx}$ \textit{vs.} $I$ with fixed $D=0.9\,$V/nm, and $0.85\leq n\leq 1.5$ (in unit of $10^{12}\,\text{cm}^{-2}$). The different line plots are vertically shifted for clarity's sake. (e) $R_{xx}$ (in log scale) \textit{vs.} inverse out-of-plane magnetic field ($1/B$) for different symmetry-breaking states. Upper panel: half-metal state with $n=3.5\times 10^{12}\,\text{cm}^{-2}$ and $D=0.75\,$V/nm; middle panel: quarter-metal state  with $n=2.6\times 10^{12}\,\text{cm}^{-2}$ and $D=0.9\,$V/nm; lower panel: Fermi-lune state exhibiting transdimensional AHE (TDAHE) with $n=1.35\times 10^{12}\,\text{cm}^{-2}$ and $D=0.9\,$V/nm.}
	\label{fig4}
\end{figure*}

Given a magnetic domain of the TOM states (say, the $\alpha^{+}$ domain), the Fermi velocities along the positive and negative crystalline $x$ direction are drastically different due to the extremely asymmetric Fermi lune structure, which would lead to significantly different resistivities for the forward-moving and backward-moving currents. Such non-reciprocal longitudinal transport behavior is directly measured in a 9-layer RMG device. In Fig.~\ref{fig4}(a), we present the longitudinal resistance ($R_{xx}$, in log scale) of the 9-layer device measured at temperature 14\,mK in the parameter space of $n$ and $D$. When $D\sim 0.75\rm{-}0.95\,$V/nm and $n\sim 0.9\rm{-}1.5\times 10^{12}\,\rm{cm}^{-2}$, the system exhibits transdimensional AHE indicating the presence of spontaneous in-plane orbital magnetization\cite{comment_exp}, and the corresponding phase boundary is marked by the red solid line in Fig.~\ref{fig4}(a). In order to demonstrate the non-reciprocal transport behavior, we fix $D=0.9\,$V/nm and present the measured longitudinal resistance $R_{xx}$ as a function of $n$ and applied current $I$, as shown in Fig.~\ref{fig4}(b). Clearly, the conductivity is highly asymmetric with respect to the direction of applied current.
In Fig.~\ref{fig4}(c) we present the line plot of $R_{xx}$ as a function of $I$ for $D=0.9\,$V/nm, with the upper and lower panels showing the data for $n=1.5\times 10^{12}\,\text{cm}^{-2}$ and $n=1.3\times 10^{12}\,\text{cm}^{-2}$, respectively. 
When $n=1.5\times 10^{12}\,\rm{cm}^{-2}$, $D=0.9\,$V/nm and $I=\pm 17\,$nA, the non-reciprocal transport signal, defined as $\delta R_{xx}/R_{xx}$, can be as large as 23\% as shown in the upper panel of Fig.~\ref{fig4}(c). Here, $\delta R_{xx}=\vert R_{xx}(+I)-R_{xx}(-I)\vert$ is the difference of resistivities measured at forward-moving and backward-moving currents with the same amplitude. This giant, intrinsic transport non-reciprocity is comparable to that reported in quantum anomalous Hall insulator Cr-doped (Bi,Sb)$_2$Te$_3$ ($\delta R_{xx}/R_{xx}\approx 26\,\%$) \cite{tokura-natnano20}, which is measured at an applied current $I=100\,\mu$A, four orders of magnitude larger than that applied to our device.  Furthermore, the microscopic mechanisms are different: in our system, the non-reciprocal longitudinal transport arises from a peculiar Fermi lune state due to interaction-driven spontaneous symmetry breaking; while in Cr-doped (Bi,Sb)$_2$Te$_3$, the large non-reciprocal transport signal is attributed to the interplay between gapless chiral edge states and Dirac surface states without the involvement of $e$-$e$ interaction effects.   

In Fig.~\ref{fig4}(d) we show line plots of $R_{xx}$ \textit{vs.} $I$ for $0.85\leq n\leq 1.5$ (in units of $10^{12}\,\rm{cm}^{-2}$). Clearly, the non-reciprocal transport behavior persists for $1.01\leq n\leq 1.5\times 10^{12}\,\rm{cm}^{-2}$. When $n=0.85$ and $0.89\times 10^{12}\,\rm{cm}^{-2}$ at $D=0.9\,$V/nm, the non-reciprocal transport behavior fades away as shown in Fig.~\ref{fig4}(d), concomitant with the vanishment of transdimensional AHE. The substantial intrinsic non-reciprocal transport signal is in perfect agreement with theoretical expectation. 
The two concomitant experimental observations, the giant non-reciprocal transport and transdimensional AHE \cite{comment_exp}, provide unambiguous evidence for the presence of Fermi lune structures and transdimensional orbital magnetism in RMG. 

Furthermore, the orbital degrees of freedom in these states show negligible response to out-of-plane magnetic fields because cyclotron orbital motion is heavily suppressed. This suppression occurs due to the pronounced anisotropy and asymmetry of the Fermi velocity distribution within the lune structure. When a vertical magnetic field is applied, the Lorentz force varies substantially in amplitude depending on the electron's direction of motion on the Fermi lune, which disrupts the establishment of closed cyclotron orbits. This is consistent with the experimental observation that no quantum oscillations are ever observed in the Fermi-lune state (exhibiting TDAHE) even when the out-of-plane magnetic field exceeds 12\,T, in contrast with other states such as the SVP state. Specifically, we select a point $n=1.35\times 10^{12}\,\text{cm}^{-2}$ and $D=0.75\,$V/nm (marked by red asterisk in Fig.~\ref{fig4}(a)) within the Fermi-lune phase, and measure $R_{xx}$ \textit{vs.} the inverse of out-of-plane magnetic field ($1/B$), as shown in the lower panel of Fig.~\ref{fig4}(e). Clearly, $R_{xx}$ remains almost as a constant even when $B$ reaches 12\,T.  In contrast, when the system is in the quarter-metal phase with $n=2.6\times 10^{12}\,\text{cm}^{-2}$ and $D=0.9\,$V/nm (marked by black square in Fig.~\ref{fig4}(a)), $R_{xx}$ exhibits clear quantum oscillations for $B\gtrapprox 2\,$T as presented in the middle panel of Fig.~\ref{fig4}(e). Similarly, when the system is in the half-metal phase with $n=3.5\times 10^{12}\,\text{cm}^{-2}$ and $D=0.75\,$V/nm (marked by black circle in Fig.~\ref{fig4}(a)), $R_{xx}$ also exhibits quantum oscillations for $B\gtrapprox 4\,$T as shown in the upper panel of Fig.~\ref{fig4}(e). 

\subsection{Mean-field phase diagrams}

\begin{figure*}[hbtp!]
	\includegraphics[width=\textwidth]{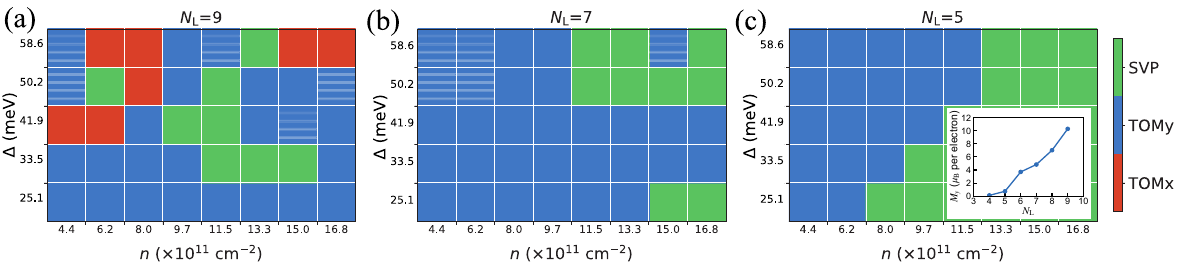}
	\caption{\textbf{Hartree-Fock phase diagrams.} 
		Hartree-Fock phase diagrams for (a) $N_{\rm{L}}=9$, (b) $N_{\rm{L}}=7$, and (c) $N_{\rm{L}}=5$. The inset in (c) presents the largest calculated in-plane orbital magnetization $M_y$ ($\mu_{\rm{B}}$ per electron) as a function of $N_{\rm{L}}$ (see text).  The blue, orange and green blocks in (a), (b) and (c) denote the TOM$_y$, TOM$_x$ and SVP states, respectively. The shaded blue blocks in (a) and (b) indicate that at these points, the TOM$_y$ is also degenerate with other non-TOM states at the mean-field level.}
	\label{fig5}
\end{figure*}

After illuminating the microscopic mechanism for the emergence of Fermi lune structures and their key physical properties, we continue to explore the full phase diagram of slightly electron-doped RMG for $4\leq N_{\rm{L}}\leq 9$.
As shown in Fig.~\ref{fig5}(a), when $N_{\rm{L}}=9$ and 7, the TOM states including both TOM$_y$ and TOM$_x$ states take most regions of the parameter space. When $N_{\rm{L}}=5$, the SVP phase starts to take a larger region, and the TOM$_y$ is more likely to emerge in the large $\Delta$ and low density regime.  The typical Fermi surfaces of TOM$_y$, TOM$_x$ and SVP states for $N_{\rm{L}}=5, 7$ and 9 are presented in Fig.~S5 of Supplementary Materials \cite{supp_info}.
The Hartree-Fock phase diagrams for $N_L=8, 6$ and 4 are also presented in Fig.~S6 of Supplementary Materials.
The transdimensional orbital magnetic properties are more prominent for thicker layers.
In particular, we find that in the TOM$_y$ state the orbital magnetization along crystalline $y$ direction $M_y$ increases with $N_{\rm{L}}$.  To illustrate this, for each $N_{\rm{L}}$ we find the TOM$_y$ state hosting the largest $M_y$ in the parameter space of ($n$, $\Delta$). Then, we plot the largest calculated $M_y$ as a function of $N_{\rm{L}}$ as shown in the inset of Fig.~\ref{fig5}(c). Clearly, it changes almost linearly with $N_{\rm{L}}$. The in-plane orbital magnetization is calculated using the numerical differentiation approach, and further details are provided in Supplementary Material\cite{supp_info}. 
We note that the in-plane orbital magnetization $\vert M_y\vert \sim \langle \hat{z} \hat{v}_x -\hat{v}_z\hat{x} \rangle$, where $\hat{x}, \hat{z}$ are position operators, and $\hat{v}_x, \hat{v}_z$ are velocity operators in $x$ and $z$ directions. The out-of-plane velocity of low-energy electrons in RMG is one order of magnitude smaller than the in-plane one, thus $\vert M_y\vert \sim \langle \hat{z} \hat{v}_x\rangle$, which is proportional to sample thickness $d$. 

\section{Discussion and conclusion}
The TOM state with Fermi-lune structure is an entirely new symmetry-breaking state of matter. Many of the physical properties are still to be explored. For instance, a metal with such peculiar Fermi lune structure may possess unusual optical properties, which are completely open questions. The collective excitations of such state is also an intriguing problem to be solved. The magnetic excitations associated with the orbital magnetization are expected to be intertwined with the charge fluctuations of the Fermi lune structure, which may give rise to intriguing collective-excitation behavior.
Moreover, there are six energetically equivalent domains for the TOM state, which would unavoidably form topological defects such as domain walls. At the intersections of the domain walls, ``$\mathbb{Z}_6$" vortices can also form as in the case of antiferromagnetic Mn$_3$Sn-class materials \cite{liu-Mn3Sn-prl-2017} and multiferroic hexagonal manganites~\cite{artyukhin-topodefect_manganite-natmat-2014,meier-topodefect_manganite-prx-2017}. The thermal phase transition from the ordered phase to disordered phase with the increase of temperature may be intimately tied with the formation and proliferation of domain walls and $\mathbb{Z}_6$ vortices, accompanied with the thermal-driven topological change of Fermi surface geometry from lune shape to double-ring shape.  Besides RMG, we would expect that the Fermi lune structure and transdimensional orbital magnetism may widely emerge in other layered van der Waals materials in the transdimensional regime. A natural extension may be multilayer transition metal dichalcogenides \cite{liu-3rtmd-science24} with slight carrier dopings.

\section*{Conflict of interest}
The authors declare that they have no conflict of interest.

\section*{Acknowledgments}
M. L., X. L. and J. L. are supported by the National Key Research and Development Program of China (grant no. 2024YFA1410400), the National Natural Science Foundation of China (grants no. U23A6002 and no. 12404221) and Shanghai Science and Technology Innovation Action Plan (grant no. 24LZ1401100). 
Q. L. and L.W. acknowledge support from the National Key Projects for Research and Development of China (Grant Nos. 2021YFA1400400, 2022YFA1204700) and Natural Science Foundation of Jiangsu Province (Grant Nos. BK20220066 and BK20233001). 
F. H. and Y. Z. acknowledge support from the National Key Research and Development Program of China (Grants No. 2022YFA1403700), NSFC 11674150, University Innovative Team in Guangdong Province 2020KCXTD001.
K.W. and T.T. acknowledge support from the JSPS KAKENHI (Grant Numbers 21H05233 and 23H02052) and World Premier International Research Center Initiative (WPI), MEXT, Japan.

\providecommand{\newblock}{}

\clearpage
\onecolumngrid
\setcounter{section}{0}
\setcounter{subsection}{0}
\setcounter{figure}{0}
\setcounter{table}{0}
\setcounter{equation}{0}
\setcounter{page}{1}
\setcounter{footnote}{0}
\renewcommand{\thefigure}{S\arabic{figure}}
\renewcommand{\thetable}{S\arabic{table}}
\renewcommand{\theequation}{S\arabic{equation}}
\renewcommand{\thepage}{S\arabic{page}}
\renewcommand{\theHsection}{supp.\arabic{section}}
\renewcommand{\theHsubsection}{supp.\arabic{section}.\arabic{subsection}}
\renewcommand{\theHfigure}{supp.\arabic{figure}}
\renewcommand{\theHtable}{supp.\arabic{table}}
\renewcommand{\theHequation}{supp.\arabic{equation}}
\renewcommand{\theHfootnote}{supp.\arabic{footnote}}
\begin{center}
{\large\bfseries Supplementary Information for ``Fermi lune and transdimensional orbital magnetism in rhombohedral multilayer graphene"}\par\medskip
Min Li$^{1,2,*}$, Qingxin Li$^{3,*}$, Xin Lu$^{1}$, Hua Fan$^{4}$, Kenji Watanabe$^{5}$, Takashi Taniguchi$^{6}$, Yue Zhao$^{4}$, Xin-Cheng Xie$^{7,8}$, Lei Wang$^{3,\dagger}$, and Jianpeng Liu$^{1,2,\ddagger}$\par\smallskip
{$^{1}$\itshape School of Physical Science and Technology, ShanghaiTech Laboratory for Topological Physics, State Key Laboratory of Quantum Functional Materials, ShanghaiTech University, Shanghai 201210, China}\par
{$^{2}$\itshape Liaoning Academy of Materials, Shenyang 110167, China}\par
{$^{3}$\itshape National Laboratory of Solid-State Microstructures, Collaborative Innovation Center of Advanced Microstructures, School of Physics, Nanjing University, Nanjing, China}\par
{$^{4}$\itshape Department of Physics, State key laboratory of quantum functional materials, Guangdong Basic Research Center of Excellence for Quantum Science, Southern University of Science and Technology, Shenzhen 518055, China}\par
{$^{5}$\itshape Research Center for Electronic and Optical Materials, National Institute for Materials Science, 1-1 Namiki, Tsukuba 305-0044, Japan}\par
{$^{6}$\itshape Research Center for Materials Nanoarchitectonics, National Institute for Materials Science, 1-1 Namiki, Tsukuba 305-0044, Japan}\par
{$^{7}$\itshape Interdisciplinary Center for Theoretical Physics and Information Sciences, Fudan University, Shanghai 200433, China}\par
{$^{8}$\itshape International Center for Quantum Materials, School of Physics, Peking University, Beijing 100871, China}\par
\end{center}
\footnotetext[1]{equal contributions}
\footnotetext[2]{leiwang@nju.edu.cn}
\footnotetext[3]{liujp@shanghaitech.edu.cn}
\setcounter{NAT@ctr}{0}


\section{Device setup and transport measurements}

The rhombohedral ennealayer graphene and hBN flakes are prepared by mechanical exfoliation of bulk crystals. Their thickness and quality were then identified by optical microscopy and atomic force microscopy. The rhombohedral domains of ennealayer graphene flakes are detected by a confocal Raman system (WITec Alpha 300R) under a 532\,nm excitation laser at room temperature. The rhombohedral domains are subsequently isolated using a anodic-oxidation-assisted atomic force microscope cutting~\cite{supp-AFM2009APL,supp-AFM2018Nanoletter}. Then we used hBN, graphite and precut ennealayer graphene pieces to assembled the multi-layer heterostructure of graphite/BN/ennealayer graphene/BN/graphite (schematically shown in Supplementary Fig.~\ref{fig_device}(a))
using dry pick-up technique with a stamp consisting of  polypropylene carbonate(PPC) film and polydimethylsiloxane(PDMS)~\cite{supp-wang2013one}. Using graphite as the gate above and below the ennealayer graphene reduces the disorder and defects introduced during evaporation compared to metal gates. Electron-beam lithography is used to write an etch mask to define the Hall-bar geometry and the electrodes. Redundant regions are etched away by CHF$_{3}$/O$_{2}$ plasma. Finally the ennealayer graphene and gates are edge-contacted by e-beam evaporating thin metal layers consisting of Cr/Pd/Au (1\,nm/15\,nm/100\,nm), as shown in Supplementary Fig.~\ref{fig_device}(b).


The transport measurements are performed in two systems, a dilution fridge with a base temperature of 15 mK and a VTI fridge down to 1.5 K, and both are with superconducting magnets. All data are taken using the standard four-terminal configuration with lock-in amplifier techniques by sourcing an AC current $I$ between 2 and 10\,nA at a frequency of 17.777\,Hz. Top and bottom graphite gates with voltages $V_{T}$ and $V_{B}$, allows us to independently tune carrier density: $n=(C_{B}V_{B}+C_{T}V_{T})/e$, and displacement field: $D=(C_{B}V_{B}-C_{T}V_{T})/2$, where $C_{T}$($C_{B}$) is top (bottom) gate capacitance. The data of longitudinal conductivity $\sigma_{xx}$ is obtained from the measured resistances by $\sigma_{xx}=\rho_{xx}/(\rho_{xx}^2+R_{xy}^2)$.

\section{Estimate of out-of-plane mean free path in rhombohedral multilayer graphene}
\label{app:mean}

From transport measurements of bulk rhombohedral graphite, one may find that the ratio between vertical and horizontal conductivities is $10^{-4}$ \cite{supp-graphite-pr53}. Assuming that such a ratio still applies to the RMG in transdimensional regime, then one can estimate the out-of-plane conductivity of RMG from the measured  in-plane conductivity $\sigma_{\perp}=10^{-4}\sigma_{\parallel}\approx 10^{-6}\,\Omega^{-1}$ when the carrier density $n\approx 10^{12}\,\text{cm}^{-2}$. Then, the relaxation time along the vertical direction $\tau_{\perp}$ can be estimated using Drude formula $\sigma_{\perp}=e^2\tau_{\perp} n/m^{*}_{\perp}$, where $m^{*}_{\perp}\approx 5 m_0$ is the out-of-plane effective mass of rhombohedral graphite ($m_0$ is electron's mass) and $e$ is elementary charge. From such information, one may estimate $\tau_{\perp}\sim 10^{-14}\,$s, and the out-of-plane Fermi velocity of RMG $v_{\perp}$ is typically one order of magnitude smaller than the in-plane Fermi velocity of graphene, with $v_{\perp}\sim 10^5\,$m/s. The out-of-plane mean free path $l_{\perp}=v_{\perp}\tau_{\perp}\approx 2\,$nm. We note that here we just make a qualitative estimate. The actual value of $l_{\perp}$ is certainly sample dependent, but it should be on the order of a few nanometers.

\section{Non-interacting continuum model for rhombohedral multilayer graphene}
The electronic structure of rhombohedral multilayer graphene (RMG) can be effectively described by the $\mathbf{k} \cdot \mathbf{p}$ Hamiltonian for a given valley $\mu$ ( where $\mu= \pm$ denotes the $\mathbf{K}/\mathbf{K}'$ valley). We begin by considering the tight-binding Hamiltonian $H_{\rm{RMG}}$ for rhombohedral multilayer graphene. The lattice vectors of graphene are defined as $\bfa_1=a_0 (1,0)$ and $\bfa_2 = a_0 (1/2,\sqrt{3}/2)$ with lattice constant $a_0=2.46$\,\AA. The corresponding reciprocal lattices are $\bfb_1= 4\pi/\sqrt{3}a_0 (\sqrt{3}/2,-1/2)$ and $\bfb_2 = 4\pi/\sqrt{3} a_0 (0,1)$. Graphene consists of two sublattices, $A$ and $B$, located at $\boldsymbol{\tau}_{1,A} =  (0,a_0/\sqrt{3})$ and $\boldsymbol{\tau}_{1,B} = (0,0)$ in Layer 1, respectively. The rhombohedral stacking of multilayer graphene is achieved by shifting each layer in the in-plane direction $(0,-a_0/\sqrt{3})$ relative to its nearest underlying layer. Consequently, each pair of neighboring graphene layers is Bernal-stacked, with the $A$ site of the top layer positioned vertically above the $B$ site of the bottom layer. In other words, for Layer $l$, the sublattice positions are given by $\boldsymbol{\tau}_{l,A} =  (0,(2-l) \times a_0/\sqrt{3})$ and $\boldsymbol{\tau}_{l,B} =  (0,(1-l) \times a_0/\sqrt{3})$, where $l=1,2,\cdots,N$. 
The tight-binding Hamiltonian for $p_z$ orbitals of carbon atoms can be written as
\begin{align}
	H_{\rm{RMG}} = \sum_{\substack{i,j,\\ l_1,l_2,\\ \alpha,\beta}} -t(\R_i + \boldsymbol{\tau}_{l_1,\alpha} + l_1 d_0  \mathbf{e}_z - \R_j - \boldsymbol{\tau}_{l_2,\beta} - l_2 d_0 \mathbf{e}_z ) \, \hcd_{l_1,\alpha} (\R_i) \hc_{l_2,\beta} (\R_j)
	\label{eq:generalTB}
\end{align}
where $i,j$ indices refer to unit cells at $\R_{i}$ and $\R_{j}$, respectively, $l_1,l_2$ are the layer indices, and $\alpha, \beta$ denote the sublattice indices. The interlayer distance is set to $d_0 = 3.35$\,\AA, and $\mathbf{e}_z$ is the unit vector perpendicular to the graphene plane. The hopping parameters are described by the Slater-Koster expressions \cite{supp-slater-lcao-pr-1954}:
\begin{subequations}
	\begin{align}
		-t(\R) &= V_{pp\pi} \left[ 1- \left(\frac{\R \cdot \mathbf{e}_z }{|\R|}\right)^2 \right] + V_{pp\sigma} \left(\frac{\R \cdot \mathbf{e}_z }{|\R|}\right)^2 , \\
		V_{pp\pi} &= V_{pp\pi}^0 \exp \left(-\frac{|\R|-a_0/\sqrt{3}}{r_0}\right) , \\
		V_{pp\sigma} &= V_{pp\sigma}^0 \exp \left(-\frac{|\R|-d_0}{r_0}\right) .
	\end{align}
\end{subequations} 
where the parameters $V_{pp\pi}^0 = -2.7$\,eV, $V_{pp\sigma}^0 = 0.48$\,eV, and $r_0 = 0.184 a_0$ are taken from prior works \cite{supp-koshino-hBN-prb14}. For this model, we consider only the interlayer hopping between two nearest graphene sheets.
Next, we perform a Fourier transform of the tight-binding Hamiltonian within the plane of the graphene layers: 
\begin{subequations}
	\begin{align}
		\hc_{l,\alpha} (\R) &= \frac{1}{\sqrt{N_c}} \sum_{\bfk} e^{i \bfk \cdot (\R + \boldsymbol{\tau}_{l,\alpha})} \hc_{l,\alpha} (\bfk) \\
		\hc_{l,\alpha} (\bfk) &= \frac{1}{\sqrt{N_c}} \sum_{\bfk} e^{-i \bfk \cdot (\R + \boldsymbol{\tau}_{l,\alpha})} \hc_{l,\alpha} (\R)
	\end{align}
\end{subequations}
where $N_c$ is the total number of unit cells. 

Expanding the tight-binding Hamiltonian around the Dirac points $\mathbf{K}/\mathbf{K}'$ (also noted as $\mathbf{K}^{+}/\mathbf{K}^{-}$, respectively, with $\mathbf{K}^{\mu} = -\mu (4\pi/3 a_0, 0)$ and $\mu=\pm 1$), we obtain the $\mathbf{k} \cdot \mathbf{p}$ Hamiltonian for valley $\mu$ as follows:
\begin{align}
	H^{0,\mu}_{\rm{RMG}}
	=    \begin{pmatrix}
		h_{\text {intra }}^{0, \mu} & \left(h_{\text {inter }}^{0, \mu}\right)^{\dagger} & 0 & 0&... & 0 \\
		h_{\text {inter }}^{0, \mu} & h_{\text {intra }}^{0, \mu} & \left(h_{\text {inter }}^{0, \mu}\right)^{\dagger} & 0&... & 0 \\
		0 & h_{\text {inter }}^{0, \mu} & h_{\text {intra }}^{0, \mu} & \left(h_{\text {inter }}^{0, \mu}\right)^{\dagger}  & ... & 0\\
		\vdots  & ...& ...& ... & ... & \vdots \\
		0 & 0 & 0 & ...& h_{\text {inter }}^{0, \mu} & h_{\text {intra }}^{0, \mu}
	\end{pmatrix}  
	\label{eq:Htot_all}
\end{align}
Here, the intralayer Hamiltonian $h^{0,\mu}_{\rm{intra}}$ is given by the $\mathbf{k} \cdot \mathbf{p}$ model for monolayer graphene:
$$
h_{\text {intra }}^{0, \mu}=-\hbar v_F^0 \mathbf{k} \cdot \boldsymbol{\sigma}_\mu,
$$
where $\hbar v_F^0 = 5.253\,\eVA$ is the non-interacting Fermi velocity of Dirac fermions extracted from the Slater-Koster tight-binding model \cite{supp-kpgra}, $\mathbf{k}$ is the wave vector expanded around the Dirac point in valley $\mu$, and $\boldsymbol{\sigma}_\mu=\left(\mu \sigma_x, \sigma_y\right)$ is the Pauli matrix defined in sublattice space. The interlayer coupling $h_{\text {inter }}^{0, \mu}$ is given by
$$
h_{\text {inter }}^{0, \mu}=\left(\begin{array}{cc}
	\hbar v_{\perp}\left(\mu k_x+i k_y\right) & t_{\perp} \\
	\hbar v_{\perp}\left(\mu k_x-i k_y\right) & \hbar v_{\perp}\left(\mu k_x+i k_y\right)
\end{array}\right),
$$
where $t_\perp = 0.34\,\eV$, $\hbar v_\perp=0.335\,\eVA$ are the interlayer hopping parameters extracted from the Slater-Koster model \cite{supp-kpgra}. In our calculations, we set up a momentum space cutoff (centered at the Dirac point) of $\pi/L_c$ with $L_c=2.25\,$nm, corresponding to an energy cutoff $E_C\sim 1\,$eV for the ennealayer model Hamiltonian given above.

The effects of an applied electric field are incorporated by introducing layer-dependent on-site energies that increase linearly with the layer index. In the experiment, the typical value of the electric field $D$ is around 1\,V/nm. With a dielectric constant $\epsilon_r=4$ and a vertical distance between two graphene sheets of $\sim d_0$, the induced energy difference due to the electric field is expected to be on the order of $\sim 80$\,meV. 
We then consider the application of an in-plane magnetic field. An in-plane field $\mathbf{B}=\left(B_x, B_y, 0\right)$ can be described by a $z$-dependent, in-plane vector potential $\mathbf{A}(\mathbf{z})=\B \times \mathbf{z}$. The effect of the in-plane field can be incorporated by replacing the $\bfk$ dependent matrix elements of the Bloch Hamiltonians \cite{supp-lee-mag-inplane}. Specifically, the matrix element $H_{l_1, l_2}(\bfk, \B)$ between Layers $l_1$ and $l_2$ ($l_1, l_2=1,2,\cdots,N$) is given by:
$$
H_{l_1, l_2}(\bfk, \B)=H_{l_1, l_2}\left(\bfk+\frac{e}{\hbar} \frac{(l_1+l_2-2) d}{2} \B \times \mathbf{e}_z\right)
$$
where $d=3.35 \, \text{\AA}$ is the interlayer distance, and $\mathbf{e}_z$ is the unit vector in the $z$ direction. When $l_1 = l_2$, this corresponds to the intralayer $h_{\text {intra }}^{0, \mu}$ part, and when $l_1 = l_2 \pm 1$ corresponds to the interlayer $h_{\text {inter }}^{0, \mu}$ part in $H^{0,\mu}_{\rm{RMG}}$.

\section{Renormalization of continuum model parameters}
In our study, we focus on the low-energy physics of rhombohedral multilayer graphene. We define a low-energy window, which includes one conduction band and one valence band per spin and per valley (namely with $n_{\text{cut}} = 1$). We classify the electrons within this low-energy window as ``low-energy electrons'' and those outside it as ``remote-band electrons''. Within this low-energy window, electron-electron interactions will be treated using a fully unrestricted self-consistent Hartree-Fock approach (see next section). However, the presence of remote-band electrons cannot be simply wiped out in the calculations because the dynamics of low-energy electrons in rhombohedral multilayer graphene are significantly influenced by remote-band electrons through long-range Coulomb interactions. As a result, the parameters of the effective low-energy single-particle Hamiltonian (Eq.~\eqref{eq:Htot_all}) are generally enhanced in magnitude compared to their non-interacting counterparts. A prominent manifestation of this enhancement is the enhancement of the Fermi velocity near the Dirac points in monolayer graphene, a phenomenon that has been extensively explored both theoretically \cite{supp-gonzalez_nuclphysb1993} and experimentally \cite{supp-elias_natphys2011}. These studies demonstrate that the Fermi velocity is notably increased due to screening effects from the filled states constituting the Dirac Fermi sea. Similarly, the interlayer coupling parameters in the low-energy model of multilayer graphene should also be renormalized accordingly \cite{supp-guo-prb24}. 

To incorporate this remote-band renormalization effects into our theoretical framework, we apply a perturbative renormalization group (RG) approach. Let us briefly sketch the process of RG flow here since a detailed derivation can be found in the Supplemental Materials of \cite{supp-guo-prb24}. First, the $e$-$e$ Coulomb interaction operator in our derivations is written as
\begin{equation}
	\hat{V}_{\text{int}} = \frac{1}{2} \int d^2 \br d^2 \br ' V_c(\br - \br ') \hat{\rho} (\br) \hat{\rho} (\br ') 
\end{equation}
where $V_c(\br) = e^2/4 \pi \epsilon_0 \epsilon_r r$ and $\hat{\rho}(\br)$ is the density operator of electrons at $\br$. The Hamiltonian Eq.~\eqref{eq:Htot_all} is defined at some high energy cutoff $\pm E_c$. Remember that the parameters should be thought of as being fixed by a measurement at $\pm E_c$ with fully empty or occupied Hilbert space and without $e$-$e$ interactions in the framework of $\bfk \cdot \p$ model. This also amounts to write $\hat{\rho}(\br)= \hat{\psi}^\dagger (\br) \hat{\psi} (\br)$ with the non-interacting field operator $\hat{\psi} (\br)$
\begin{equation}
	\hat{\psi} (\br) = \sum_{\substack{\sigma, n, 
			\bfk;\\ |\epsilon_{n,\bfk}| \leq E_c }}  \phi_{\sigma n \bfk} (\br) \hc_{\sigma n}(\bfk) 
\end{equation}
where $\phi_{\sigma n \bfk} (\br)$ is the wavefunction of an eigenstate of the non-interacting Hamiltonian Eq.~\eqref{eq:Htot_all} with energy $\epsilon_{n,\bfk}$ and its associated annihilation operator is $\hc_{\sigma n}(\bfk)$. The band index $n$ includes the bands from the two valleys $\mu=\pm 1$.

When we change the cutoff $E_c$ to a smaller one $E_c '$, these parameters are modified by $\hat{V}_{\text{int}}$, which can be treated perturbatively when $E_c '$ is much larger than any other energy scale in the system. To do so, we split the field operator $\hat{\psi} (\br) = \hat{\psi}^{<} (\br) + \hat{\psi}^{>} (\br)$ where 
\begin{align}
	\hat{\psi}^{<} (\br) &= \sum_{\substack{\sigma, n, 
			\bfk;\\ |\epsilon_{n,\bfk}| \leq E_c ' }} \phi_{\sigma n \bfk} (\br) \hc_{\sigma n}(\bfk) \\
	\hat{\psi}^{>} (\br) &= \sum_{\substack{\sigma, n, 
			\bfk;\\ E_c ' < |\epsilon_{n,\bfk}| \leq E_c }} \phi_{\sigma n \bfk} (\br) \hc_{\sigma n}(\bfk). 
\end{align}
where $\sigma$, $n$, $\bfk$ refer to spin, band index, and wavevector respectively. 

Then, we integrate out the fast modes $\hat{\psi}^{>} (\br)$ in the expansion of $\hat{\rho}(\br) \hat{\rho}(\br')$ by taking the non-interacting mean value $\langle \dots \rangle_0$ and get the expression of $\hat{V}_{\text{int}}$ at $E_c '$ as well as the quadratic correction terms to the parameters in Eq.~\eqref{eq:Htot_all}. In practice, since we need to shrink the energy window for the valence and conduction bands at the same time, this requires particle-hole symmetry \cite{supp-vafek_prl2020} such that we can always get the Coulomb interaction in the same form. In our system described by Eq.~\eqref{eq:Htot_all}, several terms break particle-hole symmetry, which actually matters knowing the fact that the nontrivial correlated states are experimentally observed only among conduction bands but absent among valence bands. However, the energy scales of the particle-hole breaking terms turn out to be negligible outside the low-energy window at which the RG flow is stopped \cite{supp-guo-prb24}. So, we can safely take advantage of this approximate particle-hole symmetry.

As this process can be done infinitesimally, we can get the following RG flow equation for the parameters in Eq.~\eqref{eq:Htot_all}:   
\begin{subequations}
	\begin{align}
		\frac{d v_F}{ d \log E_c} &= - \frac{e^2}{16 \pi \epsilon_0 \epsilon_r \hbar}\\
		\frac{d t_\perp}{d \log E_c} &= - \frac{e^2}{16 \pi \epsilon_0 \epsilon_r \hbar v_F} \\
		\frac{d v_\perp}{d \log E_c} &= 0
	\end{align}
	\label{eq:RG_flow}
\end{subequations}

Finally, after solving these differential equations above, we get
\begin{subequations}
	\begin{align}
		v_F(E_c^*)&=v_F^0 \left(1+\frac{\alpha_0}{4\epsilon_r}\log{\frac{E_C}{E_C^*}} \right)\; \label{eq:H-RG-a}\\
		t_{\perp}(E_c^*)&=t_{\perp}\,\left(1+\frac{\alpha_0}{4\epsilon_r}\log{\frac{E_C}{E_C^*}}\right)\; \label{eq:H-RG-c}\\
		v_{\perp}(E_c^*)&=v_{\perp}\;,
		\label{eq:H-RG-f}
	\end{align}
	\label{eq:H-RG}
\end{subequations}
where $\alpha_0=e^2/4\pi\epsilon_0\hbar v_F^{0}$ is the effective fine structure constant of graphene, where $\hbar$ is the reduced Planck constant and $\epsilon_0$ is the vacuum permittivity. In practice, the ratio $ L_s / n_{\text{cut}} a_0$ is used in place of $E_c / E_c^*$. Any other choice of $E_c / E_c^*$ of the same order magnitude will not affect our results since they enter in the renormalized parameter via log-scale.
\section{Hartree-Fock approximations to electron-electron interactions}
We now turn to the Coulomb interactions that impact the low-energy behavior. The electron-electron interaction operator is given by:
\begin{equation}
	\hat{V}_\text{ee}=\frac{1}{2}\int d^2 r  d^2 r' \sum _{\sigma, \sigma '} \hat{\psi}_\sigma ^{\dagger}(\br)\hat{\psi}_{\sigma '}^{\dagger}(\br ') V_c (|\br -\br '|) \hat{\psi}_{\sigma '}(\br ') \hat{\psi}_{\sigma}(\br)
	\label{eq:coulomb}
\end{equation}
where $\hat{\psi}_{\sigma}(\br)$ is the electron annihilation operator at position $\br$ with spin $\sigma$. In terms of lattice operators, this interaction can be expressed as:
\begin{equation}
	\hat{V}_\text{ee}=\frac{1}{2}\sum _{i i' j j'}\sum_{ m m'}\sum _{\alpha \alpha '\beta \beta ' }\sum _{\sigma \sigma '} \hat{c}^{\dagger}_{i, \sigma \alpha}\hat{c}^{\dagger}_{i', \sigma ' \alpha '} V^{\alpha \beta m \sigma , \alpha ' \beta ' m' \sigma '} _{ij,i'j'}\hat{c}_{j', \sigma ' m' \beta '} \hat{c}_{j, \sigma m \beta}\;,
\end{equation}
where the interaction matrix elements are:
\begin{align}
	&V^{\alpha \beta m \sigma , \alpha ' \beta ' m' \sigma '} _{ij,i'j'} \nonumber \\
	&=\int d^2 r d^2 r'  V_c (|\br -\br '|) \,\phi ^*_{\alpha} (\mathbf{r}-\mathbf{R}_i-\boldsymbol{\tau}_{\alpha})\,\phi_{m,\beta} (\mathbf{r}-\mathbf{R}_j- \boldsymbol{\tau}_{m,\beta}) \phi^*_{ \alpha  '}(\br-\mathbf{R}_i'-\boldsymbol{\tau}_{\alpha '})\phi _{m', \beta  '}(\br-\mathbf{R}_j'-\boldsymbol{\tau} _{m', \beta '}) \nonumber \\
	&\quad \times \chi ^{\dagger}_\sigma \chi ^{\dagger}_{\sigma '}\chi _{\sigma '}\chi _{\sigma} .
\end{align}
where $i, \alpha, \sigma$ refer to the atomic lattice vectors, layer/sublattice, and spin indices, respectively, $\phi$ are the Wannier functions, and $\chi$ represents the two-component spinor wave function.
To simplify the Coulomb interaction, we focus on the "density-density" interactions, which dominate in this system, and approximate the interaction matrix elements as:
$$V^{\alpha \beta m \sigma , \alpha ' \beta ' m' \sigma '} _{ij,i'j'}\approx V^{\alpha \alpha \sigma , \alpha ' \alpha ' \sigma '} _{ii,i'i'}\equiv V_{i \sigma \alpha  ,i' \sigma ' \alpha '}$$.
This leads to a simplified form of the Coulomb interaction:
\begin{align}
	\hat{V}_\text{ee}=&\frac{1}{2}\sum _{i i'}\sum _{\alpha \alpha '}\sum _{\sigma \sigma '}\hat{c}^{\dagger}_{i, \sigma \alpha}\hat{c}^{\dagger}_{i', \sigma' \alpha} V_{i\sigma \alpha, i' \sigma ' \alpha '}\hat{c}_{i', \sigma ' \alpha '}\hat{c}_{i, \sigma \alpha} \nonumber \\
	\approx&\frac{1}{2}\sum _{i \alpha \neq i' \alpha '}\sum _{\sigma \sigma '}\hat{c}^{\dagger}_{i, \sigma \alpha} \hat{c}^{\dagger}_{i', \sigma ' \alpha '} V_{i \alpha,i' \alpha '}\hat{c}_{i', \sigma ' \alpha '}\hat{c}_{i, \sigma \alpha} \nonumber 
\end{align}
Here we neglect on-site Coulomb interactions, such as the Hubbard interactions. This is because when the atomic on-site Hubbard interaction $U_0$ ($U_0\sim 1\text{-}5\,$eV) is projected to the low-energy states, it is suppressed by an order of $n\Omega_0$, where $n$ is the carrier density of the system and $\Omega_0$ is graphene's unit-cell area. Therefore, At low carrier densities, $n \sim 10^{12}\,\text{cm}^{-2}$, the effective atomic Hubbard interaction acting on the low-energy electrons becomes $\sim U_0 n\Omega_0 \sim U_0 \times 10^{-3}$. In contrast, the long-range inter-site Coulomb interactions between low-energy electrons are characterized by $\sim \frac{e^2}{4\pi\epsilon_{\rm{BN}}\epsilon_0 L_s}$, where $\epsilon_{\rm{BN}} \approx 4$ is the relative dielectric constant of the hBN substrate, and $L_s \sim \sqrt{1/\pi n}$ is the characteristic inter-electron spacing. At a density of $n \sim 10^{12}\,\text{cm}^{-2}$, the inter-electron distance is approximately $L_s \sim 5\,\text{nm}$, leading to the inter-site Coulomb interaction being much larger than the on-site Hubbard interaction. To summarize, at low electron densities (around $10^{12}\,\text{cm}^{-2}$), i.e., a few electrons per supercell, the probability of two electrons occupying the same atomic site is very low. Therefore, the dominant contribution to the Coulomb interactions between electrons arises from the long-range inter-site interactions.

To model the screening of $e$-$e$  Coulomb interactions, we adopt the double-gate screening form for Coulomb interaction. The Fourier transform of the Coulomb potential is given by:
\begin{equation}
	V(\mathbf{q})\!=\!e^2\tanh(|\mathbf{q}|d_s)/(\,2 \epsilon_{\textrm{BN}}\varepsilon_0 |\mathbf{q}|\,)
	\label{eq:V_thomasfermi}
\end{equation}
where $\mathbf{q}$ denotes wavevector and $d_s$ is the thickness of hexagonal BN substrate, which is set to 40\,nm throughout the calculations.

Since we are concerned with low-energy bands, the intersite Coulomb interactions are split into intra-valley and inter-valley terms. In this work, The intervalley term is about two orders of magnitude weaker than the intravalley one at low densities $n\sim 10^{12}\,\text{cm}^{-2}$. The intra-valley term $\hat{V}^{\text{intra}}$ can be expressed as
\begin{equation}
	\hat{V}^{\rm{intra}}=\frac{1}{2S}\sum_{\alpha\alpha '}\sum_{\mu\mu ',\sigma\sigma '}\sum_{\bk \bk ' \bq} V_(\bq)\,
	\hat{c}^{\dagger}_{\sigma \mu \alpha}(\bk+\bq) \hat{c}^{\dagger}_{\sigma' \mu ' \alpha '}(\bk ' - \bq) \hat{c}_{\sigma ' \mu ' \alpha '}(\bk ')\hat{c}_{\sigma \mu \alpha}(\bk)\;,
	\label{eq:h-intra}
\end{equation}
where $S$ is the total area of the entire system.

To facilitate the calculation of the Coulomb interaction in the band basis, we perform a transformation of the electron annihilation operators from the original basis to the band basis:
\begin{equation}
	\hat{c}_{\sigma\mu \alpha}(\bk)\equiv \hat{c}_{\sigma\mu \alpha \G}(\btk) =\sum_n C_{ \mu \alpha \mathbf{G},n}(\btk)\,\hat{c}_{\sigma \mu,n}(\btk)\;,
	\label{eq:transform}
\end{equation}
where $C_{\mu \alpha \mathbf{G},n}(\btk)$ are the expansion coefficient in the $n$-th Bloch eigenstate at $\btk$ of valley $\mu$: 
\begin{equation}
	\ket{\sigma \mu, n; \btk}=\sum_{\alpha \mathbf{G}}C_{\mu \alpha \mathbf{G},n}(\btk)\,\ket{ \sigma, \mu, \alpha, \mathbf{G}; \btk }\;.
\end{equation}
We note that the non-interacting Bloch functions are spin degenerate due to the separate spin rotational symmetry ($SU(2)\otimes SU(2)$ symmetry) of each valley. This allows us to express the intra-valley Coulomb interaction in the band basis:
\begin{align}
	\hat{V}^{\rm{intra}}&=\frac{1}{2S}\sum _{\btk \btk'\btq}\sum_{\substack{\mu\mu' \\ \sigma\sigma'}}\sum_{\substack{nm\\ n'm'}} \left(\sum _{\mathbf{Q}}\,V(\mathbf{Q}+\btq)\,\Omega^{\mu,\mu'}_{nm,n'm'}(\btk,\btk',\btq,\mathbf{Q})\right) \nonumber \\
	&\times \hat{c}^{\dagger}_{\sigma\mu,n}(\btk+\btq) \hat{c}^{\dagger}_{\sigma'\mu',n'}(\btk'-\btq) \hat{c}_{\sigma'\mu',m'}(\btk') \hat{c}_{\sigma\mu,m}(\btk)
	\label{eq:Hintra-band}
\end{align}
Here the form factor $\Omega ^{\mu,\mu'}_{nm,n'm'}$ are expressed respectively as
\begin{equation}
	\Omega ^{\mu ,\mu'}_{nm,n'm'}(\btk,\btk',\btq,\mathbf{Q})
	=\sum _{\alpha\alpha'\mathbf{G}\mathbf{G}'}C^*_{\mu \alpha\mathbf{G}+\mathbf{Q},n}(\btk+\btq) C^*_{\mu'\alpha'\mathbf{G}'-\mathbf{Q},n'}(\btk'-\btq)C_{\mu'\alpha'\mathbf{G}',m'}(\btk')C_{\mu \alpha\mathbf{G},m}(\btk).
\end{equation}

To proceed with the calculation, we apply the Hartree-Fock approximation to the Coulomb interaction, decomposing the two-particle Hamiltonian into the Hartree and Fock terms. The Hartree term is given by:
\begin{equation}
	\begin{split}
		\hat{V}_H^{\rm{intra}}=&\frac{1}{2S}\sum _{\btk \btk'}\sum _{\substack{\mu\mu'\\ \sigma\sigma'}}\sum_{\substack{nm\\ n'm'}}\left(\sum _{\mathbf{Q}} V (\mathbf{Q}) \, \Omega^{\mu ,\mu'}_{nm,n'm'}(\btk,\btk',0,\mathbf{Q})\right)\\
		&\times \left(\langle \hat{c}^{\dagger}_{\sigma\mu,n}(\btk)\hat{c}_{\sigma\mu,m}(\btk)\rangle \hat{c}^{\dagger}_{\sigma'\mu',n'}(\btk')\hat{c}_{\sigma'\mu',m'}(\btk') + \langle \hat{c}^{\dagger}_{\sigma'\mu',n'}(\btk')\hat{c}_{\sigma'\mu',m'}(\btk')\rangle \hat{c}^{\dagger}_{\sigma\mu,n}(\btk)\hat{c}_{\sigma\mu,m}(\btk)\right)
	\end{split}
	\label{eq:hartree}
\end{equation}
where the expectation values represent the self-consistent electron densities.

The Fock term is expressed as:
\begin{equation}
	\begin{split}
		\hat{V}_F^{\rm{intra}}&=-\frac{1}{2S}\sum_{\btk \btk'}\sum _{\substack{\mu\mu'\\ \sigma}} \sum_{\substack{nm\\ n'm'}} \left( \sum_{\mathbf{Q}} V (\btk'-\btk+\mathbf{Q}) \, \Omega^{\mu ,\mu'}_{nm,n'm'}(\btk, \btk ', \btk '-\btk,\mathbf{Q})\right)\\
		&\times \left(\langle \hat{c}^{\dagger}_{\sigma\mu,n}(\btk')\hat{c}_{\sigma\mu',m'}(\btk')\rangle \hat{c}^{\dagger}_{\sigma\mu',n'}(\btk)\hat{c}_{\sigma\mu,m}(\btk) + \langle \hat{c}^{\dagger}_{\sigma\mu',n'}(\btk)\hat{c}_{\sigma\mu,m}(\btk)\rangle \hat{c}^{\dagger}_{\sigma\mu,n}(\btk')\hat{c}_{\sigma\mu',m'}(\btk')\right)\;.   
	\end{split}
	\label{eq:fock}
\end{equation}

After integrating out the fast modes from the remote bands, we obtain an effective low-energy model with a cutoff $E_C^* \sim 0.3$\,eV. To handle this model, we retain one valence and one conduction band per spin and valley ($n_{\rm{cut}}$ = 1), and project the Coulomb interactions onto the non-interacting wavefunctions of the renormalized low-energy continuum model. These interactions lead to a renormalization of the continuum model parameters, while interactions within the low-energy window have not yet been included.

Then, We perform unrestricted self consistent Hartree-Fock calculations with the low-energy window, assuming 128 possible initial symmetry-breaking states characterized by the order parameters $\hat{\tau}_{a} \hat{l}_{b} \hat{\sigma}_{c} \hat{s}_{0,z}$, with $a, b, c=0, x, y, z$, where $\hat{\tau}$, $\hat{l}$, $\hat{\sigma}$ and $\hat{s}$ denote Pauli matrices in the valley, layer, sublattice and spin space, respectively, and the layer pseudospin Pauli matrix is defined
for the top and bottom layer of multilayer rhombohedral graphene. A $46\times 46$ $\mathbf{k}$ mesh is used in the Hartree-Fock calculations, with a fixed momentum space cutoff of $\pi/L_c$ with $L_c=2.25\,$nm, corresponding to an energy cutoff $E_C\sim 1\,$eV.

\section{Calculations of in-plane orbital magnetizations}
\label{app:orb}
The in-plane orbital magnetization, say along the crystalline $y$ direction, is defined as $M_y=-(e/Sd)\sum_{\bfk,n}\langle \psi_{n\bfk}\vert \hat{z} \hat{v}_x - \hat{v}_z\hat{x}\vert\psi_{n\bfk}\rangle f_{n\bfk}$. Here $\vert\psi_{n\bfk}\rangle$ is the  Bloch eigenstate at wavevector $\bfk$ with band index $n$ calculated from a given symmetr-breaking state. $f_{n\bfk}$ is the Fermi-Dirac distribution function for the corresponding Hartree-Fock Bloch band structures. $S$ is the total area and $d$ is the vertical thickness of the sample. $\hat{z}$, $\hat{x}$ are position operators, and $\hat{v}_z$ and $\hat{v}_x$ are velocity operators. Since all the symmetry-breaking ground states of interest are metallic, it would be difficult to evaluate the in-plane position operator $\hat{x}$ or $\hat{y}$ in the Bloch representation. Therefore, we use an alternative approach to calculate the in-plane orbital magnetization, which is to apply an extremely weak in-plane magnetic field, say, along $y$ direction, with $\delta B_y=0.001\,$T. The system in the TOM$_y$ state would couple to $B_y$ through the Zeeman coupling,  $\delta E_{\rm{zeem.}}=-M_y \delta B_y \Omega_0 d$ ($\Omega_0$ is the area of each unit cell).  Thus, the orbital magnetization in the TOM$_y$ state is obtained by taking the numerical differentiation of the Zeeman energy (per unit cell) with respect to $\delta B_y$: 
\begin{equation}
	M_y=-\frac{1}{\Omega_0 d}\frac{\delta E_{\rm{zeem.}}}{\delta B_y}
\end{equation}
More specifically, $\delta E_{\rm{zeem.}}=E_{\rm{tot}}(\delta B_y)-E_{\rm{tot}}(0)$ is numerically evaluated by taking the difference of the total energy  (per unit cell) when the system is orbital coupled with $\delta B_y=0.001\,$T ($E_{\rm{tot}}(\delta B_y)$) and in the absence of in-plane magnetic field ($E_{\rm{tot}}(0)$).  
More specifically, we first apply a small $\delta B_y$ to the system which is orbital coupled to the  Hartree-Fock ground state calculated under zero magnetic field. Then, we calculate  the ``one-shot" energy change of the ground state after being orbital coupled to $\delta B_y$. The orbital magnetization along the crystalline $x$ direction for the TOM$_x$ state is calculated in a similar manner. Finally, once the orbital magnetization per unit cell is obtained, we can easily convert it to the orbital moment carried by each electron given the carrier density of the system.

\section{Non-interacting fermi surfaces}
In this section, we present the non-interacting Fermi surfaces of rhombohedral multilayer graphene with layer numbers $N_{\text{L}} = 8, 7, 6$ under varying $\Delta$ and carrier density $n$, as shown in Supplementary Fig.~\ref{sup_fig1}–\ref{sup_fig3_2}. The Fermi surface undergoes significant evolution with both parameters. As the interlayer electrostatic potential energy drop $\Delta$ (induced by the vertical displacement field) increases, a Lifshitz transition occurs, marked by a topological change in the Fermi surface geometry from a circular shape to a double-ring shape. This transition also depends on the number of layers $N_{\text{L}}$. For instance, as shown in Supplementary Fig.~\ref{sup_fig1}, the Lifshitz transition occurs at $\Delta = 25.1$~meV for 8-layer graphene, while for 6-layer graphene it takes place at a higher interlayer electrostatic potential energy drop of $\Delta = 33.5$~meV. This layer dependence provides an explanation for the earlier emergence of TOM states in thicker multilayers. Additionally, increasing the carrier density leads to further evolution of the Fermi surface. In particular, when the Fermi energy crosses the Dirac point, an additional circular Fermi surface appears around it, as illustrated in Supplementary Fig.~\ref{sup_fig2}(c).

\section{Typical Fermi surfaces of interacting ground states of RMG}
Here, in Supplementary Fig.~\ref{figFS} we present typical Fermi surfaces of interacting ground states in slightly electron-doped RMG. In Supplementary Fig.~\ref{figFS}(a), we show the Fermi surfaces of typical symmetry-breaking states in 9-layer RMG. From left to right: the TOM$_y$ state, spin-valley polarized (SVP) state with double-ring shaped Fermi surface, TOM$_x$ state with twofold spin degeneracy, and TOM$_y$ state with twofold spin degeneracy and rotated by 120$^{\circ}$ compared to the left-most panel.  In Supplementary Fig.~\ref{figFS}(b), we present the typical Fermi surfaces of symmetry-breaking states in 7-layer RMG. From left to right: TOM$_y$ state with ``half-moon" shaped Fermi surface, SVP state with  circular Fermi surface, TOM$_y$ state with Fermi lune structure and SVP state with double-ring shaped Fermi surface. Finally, in Supplementary Fig.~\ref{figFS}(c), we present the Fermi surfaces of interacting ground states in 5-layer RMG. From left to right: TOM$_y$ state with half-moon shaped Fermi surface, SVP state with trigonally warped circular Fermi surface, TOM$_y$ state with Fermi lune structure and SVP state with trigonally warped circular Fermi surface. We note that the TOM$_x$ state is mostly spin degenerate, while the TOM$_y$ state may or may not have  spin degeneracy.  Here we do not discuss the spin properties of the TOM states in depth, since the spin degrees of freedom here do not contribute to orbital transport properties due to the negligible spin-orbit coupling effects.

\section{More Hartree-Fock phase diagrams of RMG}
Here we present the Hartree-Fock phase diagrams of slightly electron-doped RMG in 
the parameter space of carrier density $n$ and vertical displacement field $D$, as shown in Supplementary Fig.~\ref{fig4_sup}. In particular, the Hartree-Fock phase diagrams for the number of layers $N_L=8,  6$ and 4 are  presented in Supplementary Fig.~\ref{fig4_sup}(a), (b) and (c), respectively. Clearly, the transdimensional orbital magnetic states are more prominent for thicker layers.

\section{In-plane current distributions}

In this section, we detail the methodology for calculating inter-site tunneling currents, which include in-plane tunneling currents and interlayer tunneling currents. The time derivative of the charge density at an atomic site in layer $l$ at sublattice $\alpha$ and unit cell $\mathbf{R}$ is given by
	\begin{equation}
		\begin{aligned}
			\frac{d n_{l\alpha\mathbf{R}}}{d t} & =\operatorname{Tr}\left[\frac{d \hat{n}_{l\alpha\mathbf{R}}}{d t} \hat{\rho}\right] \\
			& =\sum_{l'\alpha'\mathbf{R}'}\left\langle\phi_{l'\alpha'}(\mathbf{R}')\right| \frac{d \hat{n}_{l\alpha\mathbf{R}}}{d t} \hat{\rho}\left|\phi_{l'\alpha'}(\mathbf{R}')\right\rangle \\
			& =\frac{i}{\hbar} \sum_{l'\alpha'\mathbf{R}'}\left\langle\phi_{l'\alpha'}(\mathbf{R}')\right|\left[\hat{H}, \hat{n}_{l\alpha\mathbf{R}}\right] \hat{\rho}\left|\phi_{l'\alpha'}(\mathbf{R}')\right\rangle \\
			& =-\frac{2}{\hbar} \operatorname{Im} \sum_{l'\alpha'\mathbf{R}'} \rho_{l\alpha, l'\alpha'}(\mathbf{R}-\mathbf{R}') H_{l'\alpha', l\alpha}(-(\mathbf{R}-\mathbf{R}')),
		\end{aligned}
	\end{equation}
	where $\{l, l'\}$ denote the layer indices and $\{\alpha, \alpha'\}$ denote the sublattice indices.  Note that $\mathbf{R}$ and $\mathbf{R}'$ are 2D in-plane lattice vectors of the 9-layer rhombohedral graphene system. The electron number operator in the Wannier basis is defined as $\hat{n}_{l\alpha}=\left|\phi_{l\alpha}\right\rangle\left\langle\phi_{l\alpha}\right|$, and the density operator in the Bloch basis is given by $\hat{\rho}=\sum_{n, \mathbf{k}}\left|\psi_{n, \mathbf{k}}\right\rangle\left\langle\psi_{n, \mathbf{k}}\right| \theta\left(E_F-E_{n, \mathbf{k}}\right)$, where $n$ is the band index and $\theta\left(E_F-E_{n, \mathbf{k}}\right)$ is the step function.

Accordingly, $\rho_{l\alpha, l'\alpha'}(\mathbf{R}-\mathbf{R}')$ can be expressed via Fourier transformation as
	\begin{equation}
		\rho_{l\alpha, l'\alpha'}(\mathbf{R}-\mathbf{R}')=\frac{1}{N} \sum_{\mathbf{k}} e^{-i \mathbf{k}\cdot(\mathbf{R}-\mathbf{R}')} \rho_{l\alpha, l'\alpha'}(\mathbf{k})
\end{equation}

and the Hamiltonian matrix element is given by

\begin{equation}
		H_{l\alpha, l'\alpha'}(-(\mathbf{R}-\mathbf{R}'))=\frac{1}{N} \sum_{\mathbf{k}} e^{i \mathbf{k}\cdot(\mathbf{R}-\mathbf{R}')} H_{l\alpha, l'\alpha'}(\mathbf{k}).
	\end{equation}
	$H_{l\alpha, l'\alpha'}(\mathbf{k})$ can be obtained from the continuum model of 9-layer rhombohedral graphene.

We define the inter-site tunneling current between the site $(l'\alpha'\mathbf{R}')$ and $(l\alpha\mathbf{R})$ as

\begin{equation}
		I_{l\alpha\mathbf{R}, l'\alpha'\mathbf{R}'}=-\frac{2}{\hbar} \operatorname{Im} \left[ \rho_{l\alpha, l'\alpha'}(\mathbf{R}-\mathbf{R}') H_{l'\alpha', l\alpha}(-(\mathbf{R}-\mathbf{R}')) \right].
\end{equation}

Using the continuum model, we can calculate all tunneling currents of TOM states, including both in-plane and out-of-plane components. The current extends across an extensive area in real spaces due to the small Fermi surface. For simplicity, we only show the in-plane currents here, which have an obvious difference in directional orientation between the TOM$_y$ and TOM$_x$ states.

Specifically, Supplementary Fig.~\ref{figs_current}(a) shows the in-plane tunneling currents for the TOM$_y$ state of 9-layer rhombohedral graphene ($\Delta=33.5\,\text{meV}$ and $n=9.7\times 10^{11}\,\text{cm}^{-2}$). These currents come from all carbon atoms (including both $A$ and $B$ sublattices) in the top layer and flow to the central $B$-sublattice atom at the origin. Blue indicates current flow from the carbon atom at coordinate $(x,y)$ to the central carbon atom, while red indicates current flow from the central carbon atom to the $(x,y)$-positioned atom in the top layer. The intensity of red or blue corresponds to the magnitude of the in-plane tunneling current. Thus, the current pattern in Supplementary Fig.~\ref{figs_current}(a) shows a net current moving left to right, along the positive crystalline x direction. Supplementary Fig.~\ref{figs_current}(b) shows the layer dependence of the average in-plane currents within a circle of radius $\sim 5~\mathrm{nm}$, with currents along the crystalline x direction (red circles) and y direction (green triangles). The top layer has a net in-plane current of about 6 to 7 nA, which decreases gradually with increasing depth into the bulk and reverses sign near the bottom layer. The bottom layer also has an average in-plane current of $\sim 6~\mathrm{nA}$, but with the opposite sign to the top layer, indicating the net current in the bottom layer flows along the negative x direction. Combined with the top layer current, this forms a closed current loop in the $x-z$ plane, leading to orbital magnetization along the crystalline y direction. In contrast, the average in-plane current along the y direction is negligible, as shown by the green triangles in Fig.~\ref{figs_current}(b), indicating negligible orbital magnetization along the crystalline x direction. We also present the real-space distribution of in-plane currents for the TOM$_x$ state in 9-layer rhombohedral graphene ($\Delta=50.2\,\text{meV}$, $n=8.0\times 10^{11}\,\text{cm}^{-2}$) in Supplementary Fig.~\ref{figs_current}(c). The in-plane tunneling currents from all carbon atoms (including both $A$ and $B$ sublattices) in the bottom layer to the central $B$-sublattice atom at the origin exhibit a net downward flow, aligned with the negative crystalline $y$-direction. Supplementary Fig.~\ref{figs_current}(d) displays the layer dependence of the average in-plane currents within a circle of radius $\sim 5~\mathrm{nm}$ for the TOM$_x$ state, which confirms the presence of orbital magnetization along the crystalline $x$-direction and negligible orbital magnetization along the crystalline $y$-direction.\\

\newpage
\begin{figure*}[hbtp!]
	\includegraphics[width=0.6 \textwidth]{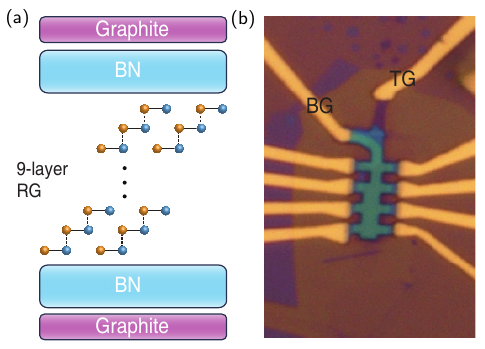}
	\caption{\textbf{Device structure and optical image.} (a) Schematic of the van der Waals stack: 9-layer rhombohedral graphene encapsulated by hBN, with graphite top and bottom gates. (b) Optical image of the device with visible gate electrodes and contacted active region.}
	\label{fig_device}
\end{figure*}

\begin{figure}[hbtp!]
	\begin{center}
		\includegraphics[width=0.7 \textwidth]{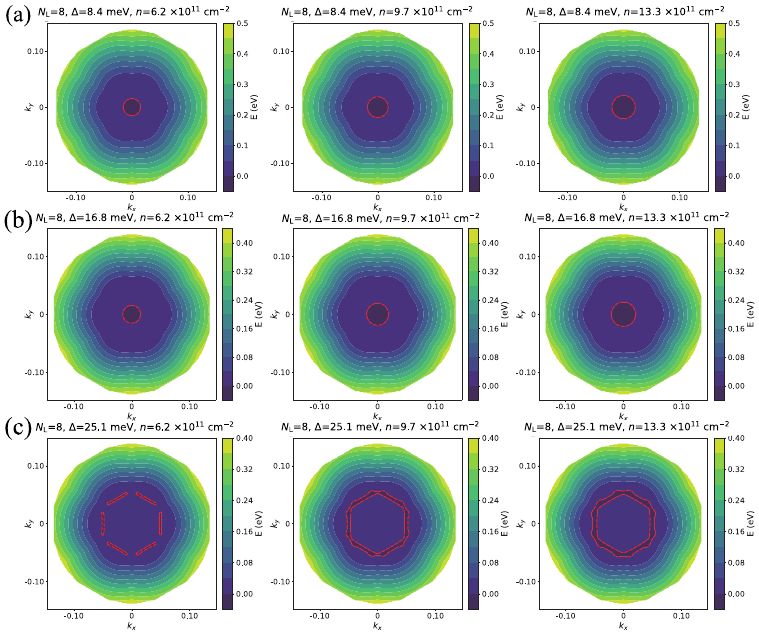}
		\caption{\textbf{Non-interacting Fermi surfaces of 8-layer RMG.} (a)–(c) Results at carrier densities of $n = 6.2 \times 10^{11}\,\text{cm}^{-2}$, $9.7 \times 10^{11}\,\text{cm}^{-2}$, and $13.3 \times 10^{11}\,\text{cm}^{-2}$, with (a) for $\Delta = 8.4\,$meV, (b) for $\Delta = 16.8\,$meV, and (c) for $\Delta = 25.1\,$meV.}
		\label{sup_fig1}
	\end{center}
\end{figure}

\begin{figure}[hbtp!]
	\begin{center}
		\includegraphics[width=0.7 \textwidth]{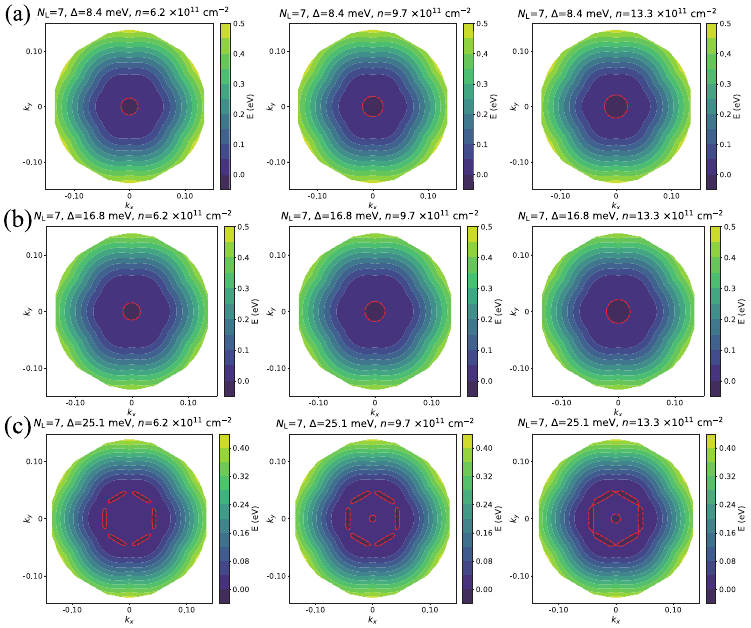}
		\caption{\textbf{Non-interacting Fermi surfaces of 7-layer RMG.} (a)–(c) Results at carrier densities of $n = 6.2 \times 10^{11}\,\text{cm}^{-2}$, $9.7 \times 10^{11}\,\text{cm}^{-2}$, and $13.3 \times 10^{11}\,\text{cm}^{-2}$, with (a) for $\Delta = 8.4\,$meV, (b) for $\Delta = 16.8\,$meV, and (c) for $\Delta = 25.1\,$meV.}
		\label{sup_fig2}
	\end{center}
\end{figure}

\begin{figure}[hbtp!]
	\begin{center}
		\includegraphics[width=0.7 \textwidth]{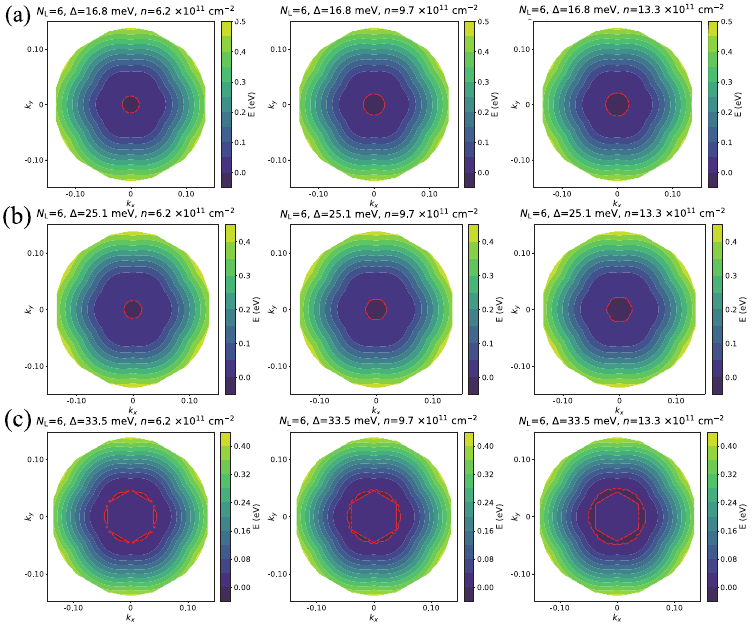}
		\caption{\textbf{Non-interacting Fermi surfaces of 6-layer RMG.} (a)–(c) Results at carrier densities of $n = 6.2 \times 10^{11}\,\text{cm}^{-2}$, $9.7 \times 10^{11}\,\text{cm}^{-2}$, and $13.3 \times 10^{11}\,\text{cm}^{-2}$, with (a) for $\Delta = 16.8\,$meV, (b) for $\Delta = 25.1\,$meV, and (c) for $\Delta = 33.5\,$meV.}
		\label{sup_fig3_2}
	\end{center}
\end{figure}

\begin{figure*}[hbtp!]
	\includegraphics[width=7in]{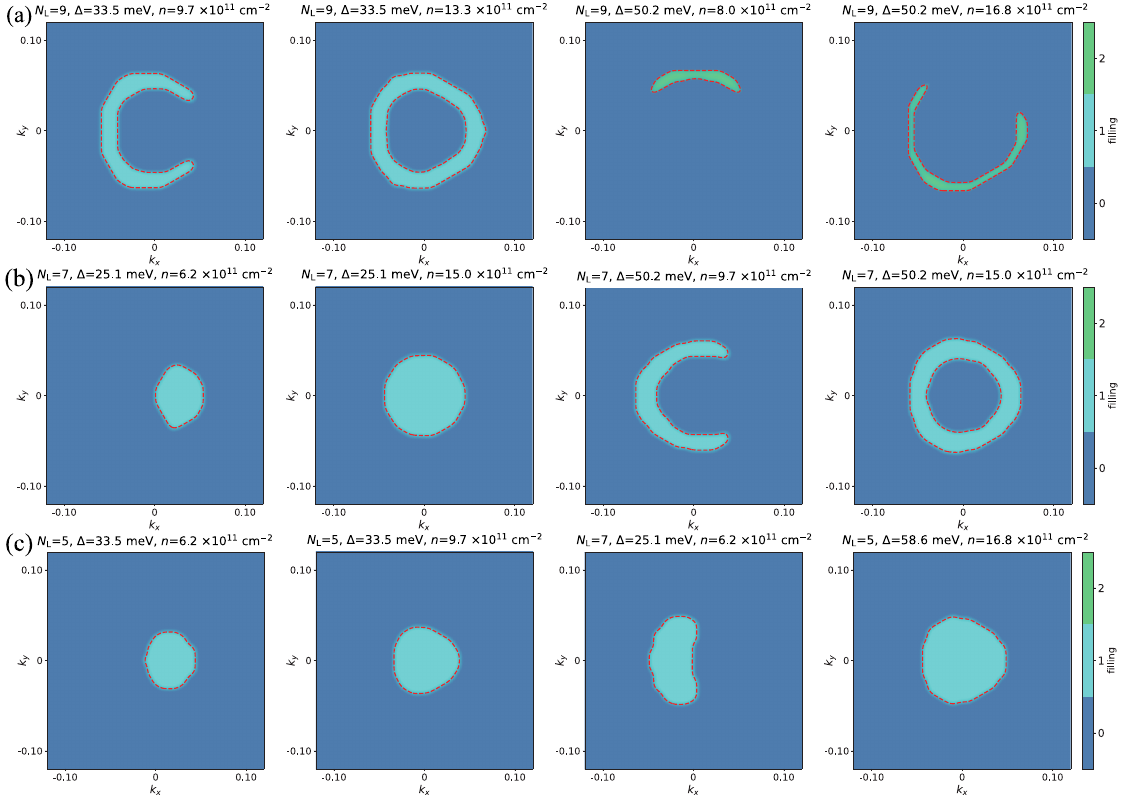}
	\caption{\textbf{Fermi surfaces of interacting ground states for RMG.} (a) For $N_{\text{L}}=9$, from left to right: $\Delta=33.5\,$meV, $n=9.7\times 10^{11}\,\text{cm}^{-2}$, TOM$_y$ state; $\Delta=33.5\,$meV, $n=1.33\times 10^{12}\,\text{cm}^{-2}$, SVP state; $\Delta=50.3\,$meV, $n=8\times 10^{11}\,\text{cm}^{-2}$, TOM$_x$ state; $\Delta=50.3\,$meV, $n=1.68\times 10^{12}\,\text{cm}^{-2}$, TOM$_y$ state. (b) For $N_{\text{L}}=7$, from left to right: $\Delta=25.1\,$meV, $n=6.2\times 10^{11}\,\text{cm}^{-2}$, TOM$_y$ state; $\Delta=25.1\,$meV, $n=1.5\times 10^{12}\,\text{cm}^{-2}$, SVP state; $\Delta=50.3\,$meV, $n=9.7\times 10^{11}\,\text{cm}^{-2}$, TOM$_y$ state; $\Delta=50.3\,$meV, $n=1.5\times 10^{12}\,\text{cm}^{-2}$, SVP state.  (c) For $N_{\text{L}}=5$, from left to right: $\Delta=33.5\,$meV, $n=6.2\times 10^{11}\,\text{cm}^{-2}$, TOM$_y$ state; $\Delta=33.5\,$meV, $n=9.7\times 10^{11}\,\text{cm}^{-2}$, SVP state; $\Delta=58.6\,$meV, $n=9.7\times 10^{11}\,\text{cm}^{-2}$, TOM$_y$ state; $\Delta=58.6\,$meV, $n=1.68\times 10^{12}\,\text{cm}^{-2}$, SVP state. The color coding indicates electron filling number at each $\mathbf{k}$ point.}
	\label{figFS}
\end{figure*}

\begin{figure*}[hbtp!]
	\includegraphics[width=7in]{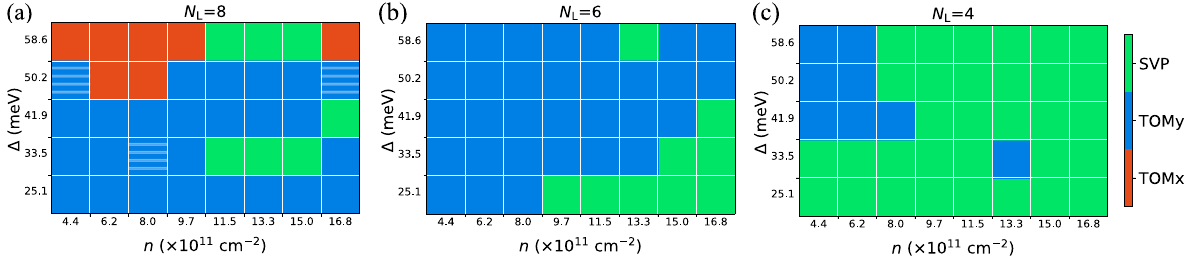}
	\caption{\textbf{Phase diagrams of slightly electron-doped RMG in the parameter space of $n$ and $\Delta$.} (a) For $N_{\text{L}} = 8$, (b) $N_{\text{L}} = 6$, (c) $N_{\text{L}} = 4$. The blue, red and green blocks denote the TOM$_y$, TOM$_x$ and SVP states, respectively. The blue blocks masked by gray horizontal lines indicate that at these points, the TOM$_y$ state is degenerate with other non-TOM states at the mean field level.}
	\label{fig4_sup}
\end{figure*}

\begin{figure*}[hbtp!]
	\includegraphics[width=7in]{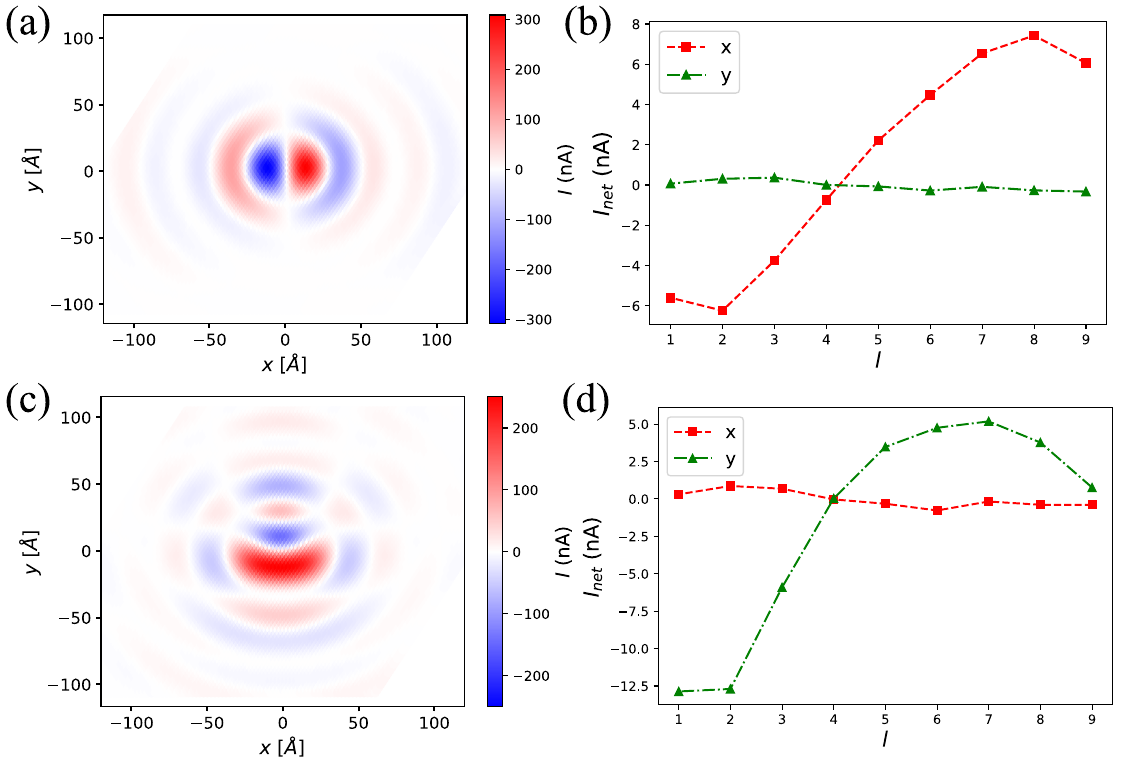}
	\caption{In-plane tunneling currentsdistribution for TOM states of 9-layer rhombohedral graphene. (a) Real-space in-plane tunneling currents from top-layer carbon atoms (A/B sublattices) to the central B-sublattice atom (origin) for the TOM$_y$ state ($\Delta=33.5\,\text{meV}$, $n=9.7\times 10^{11}\,\text{cm}^{-2}$); blue/red indicates current direction (to/from central atom), color intensity reflects current magnitude, showing net current along $+x$. (b) Layer dependence of average in-plane currents ($\sim5\,\text{nm}$ radius circle) for TOM$_y$ state; red circles ($x$-direction), green triangles ($y$-direction); closed $x$-$z$ current loop induces $y$-direction orbital magnetization, $y$-direction current is negligible. (c) Real-space in-plane tunneling currents from bottom-layer carbon atoms (A/B sublattices) to central B-sublattice atom (origin) for the TOM$_x$ state ($\Delta=50.2\,\text{meV}$, $n=8.0\times 10^{11}\,\text{cm}^{-2}$), showing net current along $-y$. (d) Layer dependence of average in-plane currents ($\sim5\,\text{nm}$ radius circle) for TOM$_x$ state, showing $x$-direction orbital magnetization and negligible $y$-direction orbital magnetization.}
	\label{figs_current}
\end{figure*}


\begin{thebibliography}{10}
\expandafter\ifx\csname url\endcsname\relax
  \def\url#1{{\tt #1}}\fi
\expandafter\ifx\csname urlprefix\endcsname\relax\def\urlprefix{URL }\fi
\providecommand{\eprint}[2][]{\url{#2}}

\bibitem{kaganov1979electron}
Kaganov M~I and Lifshits I~M 1979 {\em Soviet Physics Uspekhi\/} {\bf 22} 904

\bibitem{springford2011electrons}
Springford M, Springford M {\em et~al.\/} 2011 {\em Electrons at the Fermi
  surface\/} (Cambridge University Press)

\bibitem{nagaosa-ahe-rmp-2010}
Nagaosa N, Sinova J, Onoda S, MacDonald A~H and Ong N~P 2010 {\em Rev. Mod.
  Phys.\/} {\bf 82}(2) 1539--1592
  \urlprefix\url{https://link.aps.org/doi/10.1103/RevModPhys.82.1539}

\bibitem{nlhe-nrp21}
Du Z, Lu H~Z and Xie X 2021 {\em Nature Reviews Physics\/} {\bf 3} 744--752

\bibitem{tokura-natnano20}
Yasuda K, Morimoto T, Yoshimi R, Mogi M, Tsukazaki A, Kawamura M, Takahashi
  K~S, Kawasaki M, Nagaosa N and Tokura Y 2020 {\em Nature Nanotechnology\/}
  {\bf 15} 831--835 ISSN 1748-3395

\bibitem{li-nm24}
Zhang N~J, Lin J~X, Chichinadze D~V, Wang Y, Watanabe K, Taniguchi T, Fu L and
  Li J~I~A 2024 {\em Nature Materials\/} {\bf 23} 356--362 ISSN 1476-4660

\bibitem{zhou-halfmetal_R3G-nature-2021}
Zhou H, Xie T, Ghazaryan A, Holder T, Ehrets J~R, Spanton E~M, Taniguchi T,
  Watanabe K, Berg E, Serbyn M {\em et~al.\/} 2021 {\em Nature\/} {\bf 598}
  429--433

\bibitem{liu-symbreakR4G-natnano-2024}
Liu K, Zheng J, Sha Y, Lyu B, Li F, Park Y, Ren Y, Watanabe K, Taniguchi T, Jia
  J, Luo W, Shi Z, Jung J and Chen G 2024 {\em Nature nanotechnology\/} {\bf
  19} 188--195

\bibitem{levitov-prb23}
Dong Z, Davydova M, Ogunnaike O and Levitov L 2023 {\em Phys. Rev. B\/} {\bf
  107}(7) 075108

\bibitem{li-arxiv23}
Lin J~X, Wang Y, Zhang N~J, Watanabe K, Taniguchi T, Fu L and Li J~I~A 2023
  Spontaneous momentum polarization and diodicity in bernal bilayer graphene
  (\textit{Preprint} \eprint{2302.04261})

\bibitem{prb15_mac_currentstate}
Jung J, Polini M and MacDonald A~H 2015 {\em Phys. Rev. B\/} {\bf 91}(15)
  155423 \urlprefix\url{https://link.aps.org/doi/10.1103/PhysRevB.91.155423}

\bibitem{prb23_fs_trilayer}
Huang C, Wolf T~M~R, Qin W, Wei N, Blinov I~V and MacDonald A~H 2023 {\em Phys.
  Rev. B\/} {\bf 107}(12) L121405
  \urlprefix\url{https://link.aps.org/doi/10.1103/PhysRevB.107.L121405}

\bibitem{comment_exp}
Li Q, Fan H, Li M, Xu Y, Song J, Watanabe K, Taniguchi T, Jiang H, Xie X~C,
  Hone J, Dean C, Zhao Y, Liu J and Wang L 2026 {\em Nature\/} {\bf 653}
  384--390 ISSN 1476-4687

\bibitem{graphite-pr53}
Dutta A~K 1953 {\em Phys. Rev.\/} {\bf 90}(2) 187--192

\bibitem{supp_info}
See Supplementary Materials for: (a) details of device setup and transport
  measurements, (b) estimate of out-of-plane mean free path in rhombohedral
  multilayer graphene, (c) details of the non-interacting continuum model for
  rhombohedral multilayer graphene, (d) renormalization of low-energy model
  parameters, (e) Hartree-Fock approximations to electron-electron
  interactions, (f) calculations of in-plane orbital magnetizations, (g) more
  results of the non-interacting Fermi surfaces, (h) typical Fermi surfaces of
  the symmetry-breaking states, (i) more Hartree-Fock phase diagrams of RMG,
  (j) details of in-plane current distributions

\bibitem{delabarrera-isospinbGr-natphys-2022}
de~la Barrera S~C, Aronson S, Zheng Z, Watanabe K, Taniguchi T, Ma Q,
  Jarillo-Herrero P and Ashoori R 2022 {\em Nature Physics\/} {\bf 18} 771--775

\bibitem{zhou-isospinSCbGr-science-2022}
Zhou H, Holleis L, Saito Y, Cohen L, Huynh W, Patterson C~L, Yang F, Taniguchi
  T, Watanabe K and Young A~F 2022 {\em Science\/} {\bf 375} 774--778
  (\textit{Preprint}
  \eprint{https://www.science.org/doi/pdf/10.1126/science.abm8386})
  \urlprefix\url{https://www.science.org/doi/abs/10.1126/science.abm8386}

\bibitem{han-symbreakR5G-natnano-2024}
Han T, Lu Z, Scuri G, Sung J, Wang J, Han T, Watanabe K, Taniguchi T, Park H
  and Ju L 2024 {\em Nature Nanotechnology\/} {\bf 19} 181--187

\bibitem{zhou-trilayer-nature21}
Zhou H, Xie T, Ghazaryan A, Holder T, Ehrets J~R, Spanton E~M, Taniguchi T,
  Watanabe K, Berg E, Serbyn M {\em et~al.\/} 2021 {\em Nature\/} {\bf 598}
  429--433

\bibitem{zhou-blg-science22}
Zhou H, Holleis L, Saito Y, Cohen L, Huynh W, Patterson C~L, Yang F, Taniguchi
  T, Watanabe K and Young A~F 2022 {\em Science\/} {\bf 375} 774--778
  (\textit{Preprint}
  \eprint{https://www.science.org/doi/pdf/10.1126/science.abm8386})

\bibitem{zhangyr-blg-nature23}
Zhang Y, Polski R, Thomson A, Lantagne-Hurtubise {\'E}, Lewandowski C, Zhou H,
  Watanabe K, Taniguchi T, Alicea J and Nadj-Perge S 2023 {\em Nature\/} {\bf
  613} 268--273 ISSN 1476-4687

\bibitem{li-blg-nature24}
Li C, Xu F, Li B, Li J, Li G, Watanabe K, Taniguchi T, Tong B, Shen J, Lu L,
  Jia J, Wu F, Liu X and Li T 2024 {\em Nature\/} {\bf 631} 300--306 ISSN
  1476-4687

\bibitem{han-chiralsc-nature25}
Han T, Lu Z, Hadjri Z, Shi L, Wu Z, Xu W, Yao Y, Cotten A~A, Sharifi~Sedeh O,
  Weldeyesus H, Yang J, Seo J, Ye S, Zhou M, Liu H, Shi G, Hua Z, Watanabe K,
  Taniguchi T, Xiong P, Zumbühl D~M, Fu L and Ju L 2025 {\em Nature\/} {\bf
  643} 654--661

\bibitem{kang-rg-prl20}
Vafek O and Kang J 2020 {\em Phys. Rev. Lett.\/} {\bf 125}(25) 257602

\bibitem{guo-prb24}
Guo Z, Lu X, Xie B and Liu J 2024 {\em Phys. Rev. B\/} {\bf 110}(7) 075109
  \urlprefix\url{https://link.aps.org/doi/10.1103/PhysRevB.110.075109}

\bibitem{liu-Mn3Sn-prl-2017}
Liu J and Balents L 2017 {\em Phys. Rev. Lett.\/} {\bf 119}(8) 087202
  \urlprefix\url{https://link.aps.org/doi/10.1103/PhysRevLett.119.087202}

\bibitem{artyukhin-topodefect_manganite-natmat-2014}
Artyukhin S, Delaney K~T, Spaldin N~A and Mostovoy M 2014 {\em Nature
  materials\/} {\bf 13} 42--49

\bibitem{meier-topodefect_manganite-prx-2017}
Meier Q~N, Lilienblum M, Griffin S~M, Conder K, Pomjakushina E, Yan Z, Bourret
  E, Meier D, Lichtenberg F, Salje E~K~H, Spaldin N~A, Fiebig M and Cano A 2017
  {\em Phys. Rev. X\/} {\bf 7}(4) 041014
  \urlprefix\url{https://link.aps.org/doi/10.1103/PhysRevX.7.041014}

\bibitem{liu-3rtmd-science24}
Qin B, Ma C, Guo Q, Li X, Wei W, Ma C, Wang Q, Liu F, Zhao M, Xue G, Qi J, Wu
  M, Hong H, Du L, Zhao Q, Gao P, Wang X, Wang E, Zhang G, Liu C and Liu K 2024
  {\em Science\/} {\bf 385} 99--104 (\textit{Preprint}
  \eprint{https://www.science.org/doi/pdf/10.1126/science.ado6038})

\end{thebibliography}

\clearpage
\begin{thebibliography}{15}
\bibitem{supp-AFM2009APL}
S. Masubuchi, M. Ono, K. Yoshida, K. Hirakawa, and T. Machida, Fabrication of graphene nanoribbon by local anodic oxidation lithography using atomic force microscope, Applied Physics Letters \textbf{94} (2009).

\bibitem{supp-AFM2018Nanoletter}
H. Li, Z. Ying, B. Lyu, A. Deng, L. Wang, T. Taniguchi, K. Watanabe, and Z. Shi, Electrode-free anodic oxidation nanolithography of low-dimensional materials, Nano letters \textbf{18}, 8011 (2018).

\bibitem{supp-wang2013one}
L. Wang, I. Meric, P. Y. Huang, Q. Gao, Y. Gao, H. Tran, T. Taniguchi, K. Watanabe, L. M. Campos, D. A. Muller, J. Guo, P. Kim, J. Hone, K. L. Shepard, and C. R. Dean, One-dimensional electrical contact to a two-dimensional material, Science \textbf{342}, 614 (2013).

\bibitem{supp-graphite-pr53}
A. K. Dutta, Electrical conductivity of single crystals of graphite, Phys. Rev. \textbf{90}, 187 (1953).

\bibitem{supp-slater-lcao-pr-1954}
J. C. Slater and G. F. Koster, Simplified lcao method for the periodic potential problem, Phys. Rev. \textbf{94}, 1498 (1954).

\bibitem{supp-koshino-hBN-prb14}
P. Moon and M. Koshino, Electronic properties of graphene/hexagonal-boron-nitride moir\'{e} superlattice, Phys. Rev. B \textbf{90}, 155406 (2014).

\bibitem{supp-kpgra}
Y. Jang, Y. Park, J. Jung, and H. Min, Chirality and correlations in the spontaneous spin-valley polarization of rhombohedral multilayer graphene, Phys. Rev. B \textbf{108}, L041101 (2023).

\bibitem{supp-lee-mag-inplane}
J. Y. Lee, E. Khalaf, S. Liu, X. Liu, Z. Hao, P. Kim, and A. Vishwanath, Theory of correlated insulating behaviour and spin-triplet superconductivity in twisted double bilayer graphene, Nature communications \textbf{10}, 5333 (2019).

\bibitem{supp-gonzalez_nuclphysb1993}
J. Gonzalez, F. Guinea, and V. M. A. H., Non-Fermi liquid behavior of electrons in the half-filled honeycomb lattice (a renormalization group approach), Nuclear Physics B \textbf{424}, 595 (1994).

\bibitem{supp-elias_natphys2011}
D. C. Elias, R. Gorbachev, A. Mayorov, S. Morozov, A. Zhukov, P. Blake, L. Ponomarenko, I. Grigorieva, K. Novoselov, F. Guinea, \emph{et al.}, Dirac cones reshaped by interaction effects in suspended graphene, Nature Physics \textbf{7}, 701 (2011).

\bibitem{supp-guo-prb24}
Z. Guo, X. Lu, B. Xie, and J. Liu, Fractional chern insulator states in multilayer graphene moir\'{e} superlattices, Phys. Rev. B \textbf{110}, 075109 (2024).

\bibitem{supp-vafek_prl2020}
O. Vafek and J. Kang, Renormalization group study of hidden symmetry in twisted bilayer graphene with coulomb interactions, Phys. Rev. Lett. \textbf{125}, 257602 (2020).


\bibitem{supp-ashvin-fci-tbg-prr20}
P. J. Ledwith, G. Tarnopolsky, E. Khalaf, and A. Vishwanath, Fractional chern insulator states in twisted bilayer graphene: An analytical approach, Phys. Rev. Res. \textbf{2}, 023237 (2020).

\bibitem{supp-wang-geometry-prl21}
J. Wang, J. Cano, A. J. Millis, Z. Liu, and B. Yang, Exact landau level description of geometry and interaction in a flatband, Phys. Rev. Lett. \textbf{127}, 246403 (2021).

\bibitem{supp-regnault-fci-prx11}
N. Regnault and B. A. Bernevig, Fractional chern insulator, Phys. Rev. X \textbf{1}, 021014 (2011).
\end{thebibliography}
\end{document}